\documentclass[journal]{IEEEtran}
\usepackage{amsmath,graphicx}
\usepackage{cite}
\usepackage{mathrsfs}
\usepackage{algorithm}
\usepackage{algorithmic}
\usepackage{setspace}
\usepackage{epstopdf}
\usepackage{multirow}
\usepackage{amsfonts}
\usepackage{balance}
\usepackage[colorlinks=true, allcolors=black]{hyperref}

\usepackage{color}
\usepackage{lineno}

\begin{document}

\title{Tuning-free Plug-and-Play Hyperspectral Image Deconvolution with Deep Priors}
%\title{Real-world Color-guided Thermal Infrared Image Super-resolution: a Dataset and Benchmark}

\author{Xiuheng~Wang,~\IEEEmembership{Graduate Student Member,~IEEE,}
        Jie~Chen,~\IEEEmembership{Senior Member,~IEEE,}
        \\and C\'edric~Richard,~\IEEEmembership{Senior Member,~IEEE.}
        %and David Brie,~\IEEEmembership{Member,~IEEE.}% <-this % stops a space
\thanks{%Manuscript received September 15, 2018; revised December 23, 2018 and February 19, 2019; accepted March 7, 2019.
A preliminary version of this work has been published in the proceedings of the 2020 IEEE International Conference on Acoustics, Speech and Signal Processing (ICASSP)~\cite{wang2020learning}. Xiuheng Wang and C\'edric Richard are with Universit\'{e} C\^{o}te d'Azur, CNRS, OCA, F-06108, Nice, France (e-mail: xiuheng.wang@oca.eu, cedric.richard@unice.fr). Jie Chen is with Centre of Intelligent Acoustics and Immersive Communications at School of Marine Science and Technology, Northwestern Polytechnical University, Xi'an, 710072, China (e-mail: dr.jie.chen@ieee.org).}}
%David Brie is with Universit\'e de Lorraine, CNRS, CRAN, F-54000, Nancy, France (e-mail: david.brie@univ-lorraine.fr).

\maketitle

\begin{abstract}
Deconvolution is a widely used strategy to mitigate the blurring and noisy degradation of hyperspectral images~(HSI) generated by the acquisition devices. This issue is usually addressed by solving an ill-posed inverse problem. While investigating proper image priors can enhance the deconvolution performance, it is not trivial to handcraft a powerful regularizer and to set the regularization parameters. To address these issues, in this paper we introduce a tuning-free Plug-and-Play (PnP) algorithm for HSI deconvolution. Specifically, we use the alternating direction method of multipliers (ADMM) to decompose the optimization problem into two iterative sub-problems.  A flexible blind 3D denoising network (B3DDN) is designed to learn deep priors and to solve the denoising sub-problem with different noise levels. A measure of 3D residual whiteness is then investigated to adjust the penalty parameters when solving the quadratic sub-problems, as well as a stopping criterion. Experimental results on both simulated and real-world data with ground-truth demonstrate the superiority of the proposed method.

\end{abstract}

\begin{IEEEkeywords}
	HSI deconvolution, Plug-and-Play, tuning-free, deep learning, residual whiteness, parameter estimation.
\end{IEEEkeywords}

\section{Introduction}
Hyperspectral imaging systems simultaneously capture images of a scene over continuous narrow spectral bands ranging from ultraviolet to visible and infrared. The high spectral resolution provided by HSIs enables us to conduct analyses that cannot be performed with conventional imaging techniques.
Benefiting from abundant spectral information, hyperspectral imaging has been widely applied to applications as diverse as remote sensing \cite{bioucas2013hyperspectral} and computer vision~\cite{yan2021object}. 
%Hyperspectral imaging has been widely applied in many fields such as remote sensing \cite{bioucas2013hyperspectral}, medical science \cite{lu2014medical}, food engineering\cite{qin2009detection}, atmospheric monitoring \cite{keith2009monitoring} and document conservation~\cite{kim2011visual}. 
However, due to various physical and hardware limitations, observed HSIs are usually blurred and corrupted by noise during the acquisition process, leading to degraded performance in subsequent analyses. Thus, it is desirable to restore images by deconvolution  (inversion of the degradation process) techniques beforehand.

Multichannel images contain abundant spectral information across neighboring wavelengths, which raises the challenge of accounting for spectral correlations while ensuring spatial
consistency compared to ordinary 2D images~\cite{bongard20113d, sarder2006deconvolution}. State-of-the-art deconvolution of multichannel (multispectral) images involves Wiener filter~\cite{galatsanos1989digital,hunt1984karhunen}, Kalman filter~\cite{tekalp1990multichannel}, and regularized least-squares~\cite{galatsanos1991least}. For hyperspectral deconvolution, an adaptive 3D Wiener filter~\cite{gaucel2006adaptive} and a filter-based linear method~\cite{bongard20113d} have been used for astronomic HSIs. 2D Fast Fourier Transforms (FFTs)  and Fourier-wavelet techniques have been considered in~\cite{thiebaut2006introduction} and~\cite{neelamani2004forward} for HSI deconvolution in order to benefit from computational efficiency in Fourier and wavelet domains. In~\cite{song2019online}, an online deconvolution algorithm was devised to process HSIs sequentially collected by a push-broom device. 

Considering that deconvolution problems are usually highly ill-posed, it is strongly desirable to incorporate prior information of images to regularize the solutions. To this end, a computationally-efficient algorithm in~\cite{henrot2012fast} performs HSI deconvolution subject to positivity constraints while accounting for spatial and spectral correlations. The work in~\cite{chang2020weighted} investigates both the spatial non-local self-similarity and spectral correlations by employing low-rank tensor priors. 
Defining proper priors and designing regularizers play a key role with these methods. However it is not a trivial task to handcraft powerful regularizers, having in mind that complex regularizers may also introduce extra difficulties in solving optimization problems. Recently, benefiting from the variable splitting principle, various PnP methods have been proposed recently. They consist of plugging image denoising modules in optimization modules to solve inverse problems. We shall now outline the main principles of the PnP framework.

Consider the general inverse problem consisting of minimizing the following objective function:
\begin{equation}\label{eq.lms_model}
  \mathbf{\hat{x}}=\arg \min_{\mathbf{x}}\mathcal{D}(\mathbf{x})+\lambda \mathcal{R}(\mathbf{x}),
\end{equation}
where $\mathbf{x}$ is the unknown variable to be estimated, $\mathcal{D}(\mathbf{x})$ is the data fidelity term that ensures the consistency between the reconstructed and observed signals, and $\mathcal{R}(\mathbf{x})$ is a regularizer that enforces desirable properties of the solution with $\lambda \ge 0$ the regularization parameter.
With the ADMM \cite{boyd2011distributed} or the half quadratic splitting method~\cite{geman1995nonlinear}, the optimization problem~\eqref{eq.lms_model} can be solved in $K$ iterations consisting of  two key operations:
% \begin{align}
%   \label{eqn1}&\mathbf{\hat{x}}_{k+1} = \arg \min_{\mathbf{x}}\mathcal{D}(\mathbf{x})
%   +\frac{\rho_k}{2}\|\mathbf{x} - \mathbf{\hat{v}}_k\|_{2}^{2}; \\
%   \label{eqn2}&\mathbf{\hat{v}}_{k+1} =  \textit{Denoiser}(\mathbf{\hat{x}}_{k+1}, \sigma_k);
% \end{align}
\begin{align}
  \label{eqn1}&\mathbf{\hat{x}} = \arg \min_{\mathbf{x}}\mathcal{D}(\mathbf{x})
  +\frac{\rho}{2}\|\mathbf{x} - \mathbf{\hat{v}}\|_{2}^{2}; \\
  \label{eqn2}&\mathbf{\hat{v}} =  \texttt{Denoiser}(\mathbf{\hat{x}}, \sigma);
\end{align}
where $\rho$ is the penalty parameter, and $\texttt{Denoiser}(\cdot)$ represents a denoising operator with $\sigma = \sqrt{{\lambda}/{\rho}}$ the denoising strength. Conversely, this formulation can also implicitly define $\mathcal{R}(\cdot)$ when plugging an arbitrary denoising operator. This allows to benefit from the merits of deep learning and optimization methods~\cite{chen2022integration}, and to eliminate the need for expensive network retraining whenever the inverse problem changes~\cite{wei2020tuning}. Applications include {magnetic resonance imaging (MRI)} reconstruction~\cite{venkatakrishnan2013plug, wei2020tuning}, 2D image restoration~\cite{brifman2016turning, zhang2017learning, chen2020learning, zhang2021plug} and hyperspectral unmixing~\cite{wang2020hyperspectral, zhao2021plug}. Despite its effectiveness, this strategy has not yet been employed in HSI deconvolution problems, though similar difficulties of designing regularizers are encountered there. 

Regardless of whether the regularizers are manually designed or implicitly learned as in recent PnP algorithms, it is desirable to select the regularization parameters properly to balance the contribution of prior information and observations. 
Classic parameter estimation methods used with handcrafted regularizers include the discrepancy principle (DP)~\cite{thompson1991study}, the L-curve~\cite{hansen1992analysis, vogel1996non}, the generalized cross-validation (GCV)~\cite{golub1979generalized, reeves1994optimal}, and Stein’s unbiased risk estimate (SURE)~\cite{stein1981estimation, van2011nonlocal}. Recently, the authors of~\cite{song2016regularization} proposed the maximum curvature criterion and the minimum distance criterion (MDC) on the response surface to estimate the regularization parameters in a non-negative HSI deconvolution problem~\cite{henrot2012fast}. The MDC has been extended to HSI super-resolution by considering a deep prior regularizer in~\cite{wang2021hyperspectral}. By defining and maximizing some whiteness measures of residual images, the authors of~\cite{almeida2013parameter} proposed a 2D image deblurring method with objective criteria for adjusting the regularization parameter as well as the stopping criterion. In~\cite{lanza2020residual}, an exact residual whiteness principle has been proposed for generalized Tikhonov-regularized 2D image restoration. However, a specific-designed criteria for 3D images, such as HSIs, is still missing.

Compared to handcrafted regularizers, implicit regularizers in PnP algorithms introduces extra challenges that need to be addressed for devising an automatic regularization parameter estimation strategy. In the PnP framework, $\lambda$ is reparameterized by a series of internal parameters, including the penalty parameter $\rho$, the denoising strength $\sigma$, and the number of iterations $K$ (related to stopping criteria). In~\cite{brifman2016turning, wang2020hyperspectral, zhao2021plug}, a constant scaling factor is used to increase $\rho$ linearly as iterations proceed. In~\cite{zhang2017learning}, $\sigma$ is exponentially decayed in sequential denoising sub-problems. Nevertheless, the selected parameters in all these handcrafted criteria may lead to sub-optimal performance since the internal parameters may not change monotonically. To address this issue, the methods in~\cite{chen2020learning, zhang2021plug} consist of training a blind denoising network to estimate $\sigma$ automatically. The work in~\cite{chen2020learning} considers a fixed $\rho$ while the approach in~\cite{zhang2021plug} considers a fixed $\lambda$. Unlike these semi-automated approaches, deep reinforcement learning is used in~\cite{wei2020tuning} to determine all the internal parameters, leading to good convergence behavior and performance. 

In this paper, we introduce a fully automatic PnP hyperspectral deconvolution method that uses spectral-spatial priors learned from data by a deep neural network. The HSI deconvolution problem is addressed with an ADMM algorithm. In order to avoid manually selecting the regularization parameters, we define a non-negative scalar measure of whiteness for 3D residual images, which cooperates with a blind deep denoiser to adaptively adjust all the internal parameters. The contributions of this work are summarized as follows:
\begin{itemize}
	\item We propose a PnP framework for
	hyperspectral deconvolution. Based on the ADMM algorithm, the optimization problem is split into two sub-problems, a simple quadratic sub-problem and a 3D-image denoising sub-problem. 
	\item A blind deep denoiser B3DDN is designed and plugged into the proposed framework. This denoising operator learns both spatial context and spectral attributes of HSIs, bypassing the difficulty in designing regularizers. After training with simulated data, the flexibility of the B3DDN allows it to learn, without any extra training, the priors for real-world images even with a distinct number of spectral channels.
	\item The proposed PnP framework is designed in a completely turning-free manner. Specifically, the penalty parameters are determined automatically by solving a scalar optimization problem while the denoising strengths are implicitly learned by the B3DDN. A stopping criterion for the iterative process is also provided.
	\item An HSI dataset containing six blurring and clear image pairs captured in indoor and outdoor scenes is provided with this work. This dataset allows us to show that our method is applicable with real-world scenarios. It also provides a benchmark for future research works in hyperspectral deconvolution.
\end{itemize}

The paper is organized as follows. In Section~\ref{sec:problem}, HSI deconvolution is formulated as a linear inverse problem. Section~\ref{sec:method} introduces the proposed tuning-free deconvolution method based on the PnP framework with learned deep priors. In
Section~\ref{sec:experiment}, experiments with simulated and real-world data are conducted and analyzed. Section~\ref{sec:conclusion} concludes this paper.

\label{method}
\begin{figure*}[!t]
	\centering
	\includegraphics[scale=0.52]{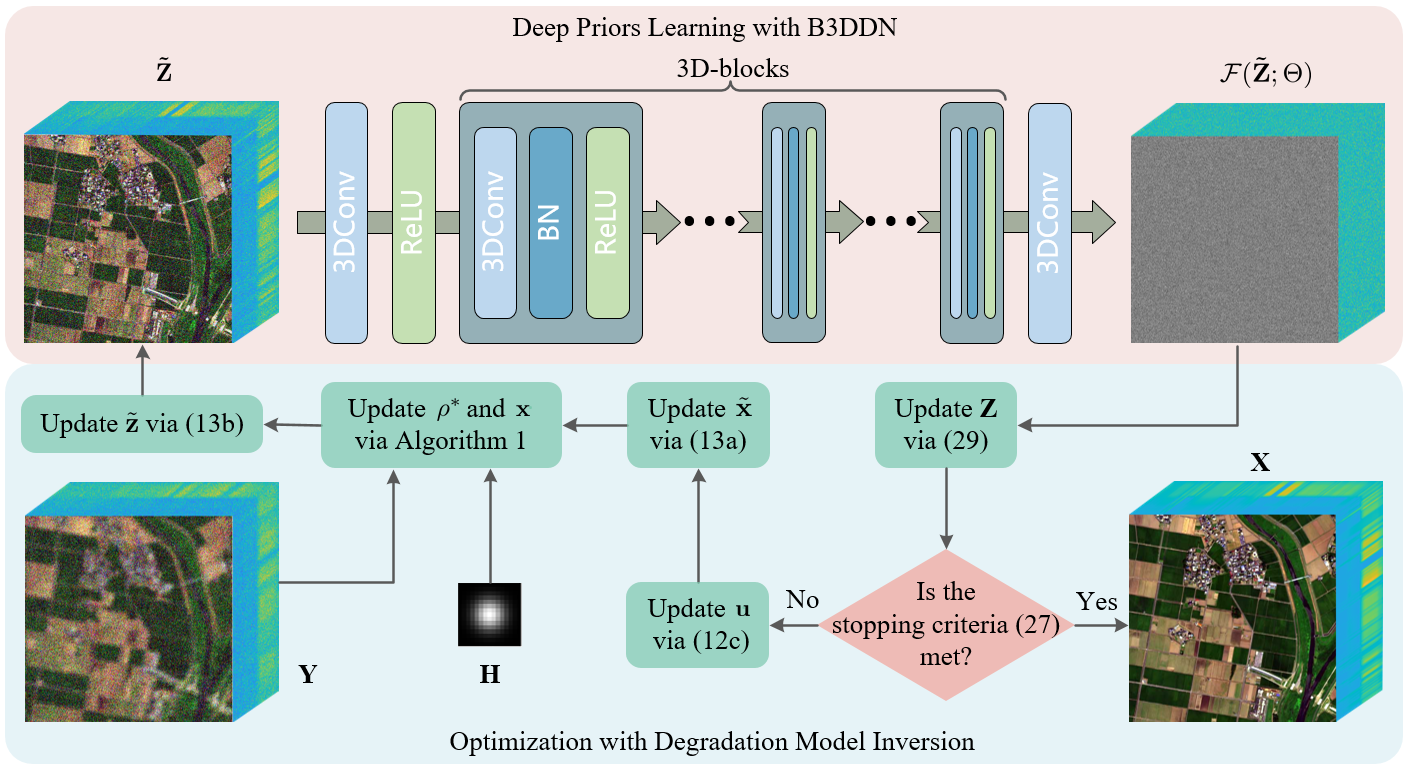}
	\caption{Architecture of the proposed tuning-free scheme for hyperspectral image deconvolution. (top) Network structure of the B3DDN. (bottom) Numerical optimization steps in the ADMM framework.}
	\label{fig:net}
\end{figure*}

\section{Problem formulation}
\label{sec:problem}
We denote a degraded HSI and its latent clean counterpart by ${\bf Y} \in \mathbb{R}^{N\times P\times Q}$ and ${\bf X} \in \mathbb{R}^{N\times P\times Q}$ respectively, where $P$, $Q$, and $N$ are the numbers of rows, columns and spectral bands of the image. Using lexicographical order, ${\bf Y}$ and ${\bf X}$ can be reshaped into vectors ${\bf y}\in \mathbb{R}^{NPQ\times 1}$ and ${\bf x} \in \mathbb{R}^{NPQ\times 1}$, respectively. The degraded image and the clean image at the $i$-th spectral band are denoted by ${\bf Y}_i \in \mathbb{R}^{P\times Q}$ and ${\bf X}_i \in \mathbb{R}^{P\times Q}$.  For ease of mathematical formulation, the columns of ${\bf Y}_i$ and ${\bf X}_i$ are stacked to form vectors ${\bf y}_i \in \mathbb{R}^{PQ\times 1}$ and ${\bf x}_i \in \mathbb{R}^{PQ\times 1}$. ${\bf x}$ and ${\bf y}$ are vectors obtained by stacking vectors ${\bf x}_i$ and ${\bf y}_i$ ($1\le i\le N$), respectively. This notation system also works for other images.
%Note that we use similar notations for other images.

For the $i$-th channel, ${\bf Y}_i$ is generated from ${\bf X}_i$ according to the following 2D degradation model:
\begin{equation}
\label{eq:convolution}
{\bf Y}_i = \mathcal{H}_i * {\bf X}_i + {\bf N}_i
\end{equation}
where $\mathcal{H}_i$ is the convolution kernel, possibly containing null entries, of size $P \times Q$ encoding the Point Spread Function (PSF) of the $i$-th channel:
\begin{equation}
\label{eq:psf}
\mathcal{H}_i  = \left(\begin{matrix}
\mathcal{H}_{11} &\cdots &\mathcal{H}_{1Q}\\
\vdots &\ddots &\vdots\\
\mathcal{H}_{P1} &\cdots &\mathcal{H}_{PQ}
\end{matrix}\right),
\end{equation}
Operator $*$ denotes the discrete 2D convolution performed in the image domain, {and $ {\bf N}_i$ is an additive independent and identically distributed (i.i.d.) Gaussian noise with standard deviation $\sigma$.} Following~\cite{henrot2012fast}, model~\eqref{eq:psf} can be written as: 
\begin{equation}
\label{eq:linear}
{\bf y}_i = {\bf H}_i{\bf x}_i + {\bf n}_i
\end{equation}
where ${\bf H}_i$ is a $PQ\times PQ$ block-Toeplitz matrix with $P\times Q$ Toeplitz blocks. Imposing periodic boundary conditions on $\mathcal{H}_i$, ${\bf H}_i$ can be rewritten as a block circulant matrix with circulant blocks, a structure denoted as circulant-block-circulant (CBC). This property allows us to design a Fourier domain implementation for solving the least square problem in Section~\ref{ssec:admm}. 
%It is well-known that CBC matrices can be diagonalized by the 2D discrete Fourier transform (DFT), this property is used in subsection~\ref{ssec:admm}.

Assuming that the convolution is separable and the noise variance is independent over spectral bands, the hyperspectral degradation model can be written as:
\begin{equation}
    \label{eq:3d}
    {\bf y} = \mathbf{H}{\bf x} + {\bf n}
\end{equation}
where $\bf H$ is a block-diagonal matrix of size $NPQ\times NPQ$:
\begin{equation}
\label{eq:H}
\mathbf{H}  = \left[\begin{matrix}
\mathbf{H}_{1} & \bf 0 &\cdots &\bf 0 \\
\bf 0  &\ddots &\ddots &\vdots\\
\vdots &\ddots &\ddots &\bf 0\\
\bf 0 & \cdots & \bf 0 & \mathbf{H}_{N}
\end{matrix}\right].
\end{equation}
The problem in HSI deconvolution is formulated as an inverse problem, where ${\bf x}$ is estimated by seeking the minimum of the following objective function:
\begin{equation}
\label{eq:objective}
\hat {\bf x} = \arg\mathop{\min}_{{\bf x}}\,\frac{1}{2}\|{\bf y}-{\bf H}{\bf x}\|^{2} + \lambda\Phi (\bf{x})
\end{equation}
where  the first squared-error term $\frac{1}{2}\|{\bf y}-{\bf H}{\bf x}\|^{2}$ is the data fidelity term, and $\Phi (\bf{x})$ is the regularizer.
%In inverse imaging problems, spatial and spectral correlations are often encoded in~$\Phi (\bf{x})$ and $\lambda$ controls the trade-off between degradation model inversion and prior information.

\section{Proposed method}
\label{sec:method}
Designing an effective regularizer $\Phi(\bf{x})$ along with an efficient solving method is not trivial. Meanwhile, it is cumbersome to fine-tune the hyperparameter $\lambda$ to balance the contribution of $\Phi(\bf{x})$ for different images. To tackle these issues, we propose to learn priors from hyperspectral data and incorporate it into the model-based optimization to tackle the regularized inverse problem in~\eqref{eq:objective}. More specifically, using the variable splitting technique, we transform problem~\eqref{eq:objective} into two sub-problems, namely, a simple quadratic problem with a penalty parameter and a 3D-image denoising problem with a certain denoising strength. These sub-problems are iteratively solved, using a linear method and a blind deep neural network, respectively, until the convergence criterion is met. In this procedure, the penalty parameter is automatically estimated while the denoising strength is implicitly learned. Finally, the algorithm is automatically terminated by stopping criteria. Our tuning-free HSI deconvolution scheme is illustrated in Fig.~\ref{fig:net}. 

\subsection{Variable splitting based on the ADMM}
\label{ssec:admm}

The ADMM is adopted to decouple the data fidelity term and the regularization term in~\eqref{eq:objective}. By introducing an auxiliary variable $\bf z$, problem~\eqref{eq:objective}  can be written in the equivalent form:
\begin{equation}\label{eq:objective2}
\begin{split}
\hat {\bf x} = &\arg\mathop{\min}_{{\bf x}}\frac{1}{2}\|{\bf y}-{\bf H}{\bf x}\|^{2} + \lambda \Phi ({\bf z}), \quad {\rm s.t.} \quad {\bf z} = {\bf x}  .
\end{split}
\end{equation}
The associated augmented Lagrangian function is given by
\begin{equation}
\label{eq:Lag}
\begin{aligned}
\mathcal{L}_{\rho}({\bf x,z,v}) = & \arg\mathop{\min}_{{\bf x}}\,\frac{1}{2}\|{\bf y}-{\bf H}{\bf x}\|^{2} + \lambda \Phi ({\bf z})\\
& + {\bf v}^T({\bf x-z})+\frac{\rho}{2}\|{\bf x-z}\|^{2}
\end{aligned}
\end{equation}
with $\bf v$ the dual variable, and $\rho > 0$ the penalty parameter. Scaling $\bf v$ as ${\bf u}=\frac{1}{\rho}{\bf v}$, problem \eqref{eq:Lag} can be iteratively solved by repeating the following successive steps:
\begin{subequations}
	\begin{align}
	\label{eq:stepa} {\bf x}_{k+1} =& \arg\mathop{\min}_{{\bf x}}\,\frac{1}{2}\|{\bf y}-{\bf H}{\bf x}\|^{2}+\frac{\rho_k}{2}\|{\bf x}-{\tilde{\bf x}}_{k}\|^{2}\\
	\label{eq:stepb} {\bf z}_{k+1} =& \arg\mathop{\min}_{{\bf z}}\,\lambda \Phi ({\bf z})+\frac{\rho_k}{2}\|{\tilde{\bf z}}_{k}-{\bf z}\|^{2}\\
	{\bf u}_{k+1} =&\, {\bf u}_k+{\bf x}_{k+1}-{\bf z}_{k+1}
	\end{align}
\end{subequations}
where 
\begin{subequations}
	\begin{align}
	\label{eq:stepc} {\tilde{\bf x}}_{k} &= {\bf z}_k - {\bf u}_k\\
	\label{eq:stepd} {\tilde{\bf z}}_{k} &= {\bf x}_{k+1} +{\bf u}_k
	\end{align}
\end{subequations}
and $\rho_k$ denotes the penalty parameter at the $k$-th iteration. In this way, the data fidelity term and the regularization term in~\eqref{eq:objective} are decoupled into two sub-problems,~\eqref{eq:stepa} and~\eqref{eq:stepb}. Sub-problem~\eqref{eq:stepa} is a least square problem that can be solved analytically as follows:
\begin{equation}\label{eq:x_solution}
{\bf x}_{k+1} = ({\bf H}^{T}{\bf H}+\rho_k {\bf I})^{-1}({\bf H}^{T}{\bf y}+\rho_k{\bf \tilde{\bf x}}_{k})
\end{equation}
% {Note this step is actually computed with efficiency in the Fourier domain:} 
% \begin{equation}\label{eq:x_solution_fft}
% {\bf x}_{k+1} = \mathcal{F}^{-1}\left(\frac{\mathcal{F}^*(\bf H)\mathcal{F}(\bf y)+\rho_k\mathcal{F}({\bf \tilde{\bf x}}_{k})}{\mathcal{F}^*(\bf H)\mathcal{F}(H)+\rho_k\mathcal{F}(\bf I)}\right)
% \end{equation}
Subproblem~\eqref{eq:stepb} can be reformulated as:
\begin{equation}\label{eq:denoise}
{\bf z}_{k+1} = \arg\mathop{\min}_{{\bf z}}\,\frac{1}{2\sigma_k^2}\|{\tilde{\bf z}}_{k}-{\bf z}\|^{2}+\Phi ({\bf z})
\end{equation}
where $\sigma_k = \sqrt{{\lambda}/{\rho_k}}$. 

From a Bayesian perspective\footnote{{Considering a degradation model ${\tilde{\bf z}}_{k} = {\bf z} + {\bf n}_k$ where ${\bf n}_k$ is Gaussian noise with standard deviation $\sigma_k$. 
The denoising problem can be formulated as the recovery of the posterior probability distribution function (PDF) $p({\bf z}|{\tilde{\bf z}}_{k})$. Using the Bayes theorem, this PDF can be written as: $p({\bf z}|{\tilde{\bf z}}_{k})\propto p({\tilde{\bf z}}_{k}|{\bf z})p({\bf z})$ where $p({\bf z})$ is the prior probability distribution of ${\bf z}$ and $\propto$ means ``proportional to".  Finally, the log-posterior distribution can be written as $-\mathrm{log}\,p({\bf z}|{\tilde{\bf z}}_{k}) = \frac{1}{2\sigma_k^2}\|{\tilde{\bf z}}_{k}-{\bf z}\|^{2} + \mathrm{log}\, p({\bf z}) + C$ where $C$ is a constant. By rewriting $\mathrm{log}\, p({\bf z})$ as $\Phi ({\bf z})$, estimating ${\bf z}$ in the sense of the maximum a posterior principle leads to the optimization problem in~(15).}}, \eqref{eq:denoise} can be considered as a denoising problem, removing Gaussian noise with noise-level $\sigma_k$ from the noisy HSI $\tilde {\bf z}_{k}$ to obtain the clean HSI ${\bf z}_{k+1}$. In other words, a denoising operator can be used for neglecting the design of the regularization term $\Phi (\bf{x})$. %Note that the denoising procedure is actually performed in the 3D image domain.

\subsection{Estimating parameters via 3D residual whiteness}

\label{ssec:subhead_2}
In most real-world applications, no ground-truth information is available for fine-tuning the algorithm parameters or terminating the optimization at a proper iteration. To tackle this issue,  a measure of residual whiteness of 3D images is defined in this subsection, and the optimal value of $\rho_k$ at each iteration, as well as the number of iterations, can be determined with the help of this measure. To be specific, we propose to evaluate the optimal $\rho_k^{*}$ in~\eqref{eq:stepa} by solving a scalar optimization problem. The stopping criterion then consists of comparing this 3D whiteness measure between two iterations.

\subsubsection{Measure of 3D residual whiteness}% The parameter estimation relies on the measure of residual whiteness. Let us consider the residual image $\mathbf{r}\in \mathbb{R}^{L}$ with $L = NPQ$ with its original 3D matrix ${\bf R} \in \mathbb{R}^{N\times P\times Q}$. The $\textit{auto-correlation}$ of $\bf R$ is defined as
% \begin{equation}\label{eq:ac}
% {\bf A}_{\bf rr} = \frac{1}{L}({\bf R} * {\bf R})
% \end{equation}
% with the $\textit{sample auto-correlation}$ at the indexes $(n, p, q)$ being
% \begin{equation}\label{eq:sac}
% {\bf A}_{\bf rr}(n,p,q) = \frac{1}{L} \sum_{m,i,j}{\bf R}(n,p,q){\bf R}(m-n,i-p,j-q)
% \end{equation}
% where $*$ is the 3D discrete correlation and $1\le m\le N, 1\le i \le P, 1\le j \le Q$. 
We define the residual image $\mathbf{r}_{k+1}\in \mathbb{R}^{L}$ with $L = NPQ$ by:
\begin{equation}
    {\mathbf{r}}_{k+1} = {\bf H}{\mathbf{x}}_{k+1} - {\bf y}
\end{equation}
with its equivalent 3D image matrix denoted by ${\bf R}_{k+1} \in \mathbb{R}^{N\times P\times Q}$. The $\textit{auto-correlation}$ of $\bf R_{k+1}$ is defined as:
\begin{equation}\label{eq:ac}
{\bf A}_{{\bf R}_{k+1}} = \frac{1}{L}({\bf R}_{k+1} * {\bf R}_{k+1})
\end{equation}
where $*$ denotes the 3D discrete correlation. The sample auto-correlation at indexes $(n, p, q)$ is given by:
\begin{equation}\label{eq:sac}
{\bf A}_{{\bf R}_{k+1}}(n,p,q)\! =\! \frac{1}{L} \sum_{m,i,j}{\bf R}_{k+1}(n,p,q){\bf R}_{k+1}(m-n,i-p,j-q)
\end{equation}
with $1\le m\le N$, $1\le i \le P$, $1\le j \le Q$.
When the residual is close to the modeling error $\mathbf{n}$, i.e., a white Gaussian noise, ${\bf A}_{{\bf R}_{k+1}}(n,p,q)$ satisfies the following asymptotic property:
\begin{equation}\label{eq:ac_pro}
\lim_{L\to\infty} {\bf A}_{{\bf R}_{k+1}}(n,p,q) \approx \begin{cases}
\ \sigma^2 \ &\text{if}\ (n,p,q) = (0,0,0)\\
\ 0 \ &\text{if}\ (n,p,q) \ne (0,0,0)
\end{cases}
\end{equation}
 The size $L$ of hyperspectral images is usually large (between $10^6$ and $10^8$), so that we can assume that the sample auto-correlation at all indexes $(n,p,q) \ne (0,0,0)$ is close to zero. %More precisely, it has been proved in~\cite{lanza2020residual} that all these quantities can be regarded as random variables approximately in a Gaussian distribution with zero mean and stand deviation $\sigma_a = \sigma^2 / NPQ$. 
This assumption is based on the following result of the Gaussian process $\bf n$ with its equivalent 3D image matrix denoted by ${\bf N} \in \mathbb{R}^{N\times P\times Q}$ and sample auto-correlation ${\bf A}_{{\bf N}_{k+1}}(n,p,q)$ defined by replacing $\bf R$ as $\bf N$ in~\eqref{eq:sac}.

{\bf Theorem 1.} \textit{If $\bf n$ has a finite variance $\sigma$ and $L$ tends to $\infty$, any ${\bf A}_{{\bf N}_{k+1}}(n,p,q)$ with $(n,p,q) \ne (0,0,0)$ is asymptotically uncorrelated and Gaussian-distributed with zero mean and stand deviation $\sigma_a = \sigma^2 / L$}

\textit{Proof.} The proof follows directly by applying Proposition~1 of~\cite{lanza2018whiteness} to the 3D domain.

The rational behind imposing residual whiteness is to estimate parameters by constraining the residual auto-correlation at non-zero indexes to be small. To make this measure independent from $\sigma$, inspired by~\cite{lanza2020residual}, we consider the \textit{normalized auto-correlation} defined as follows:
\begin{equation}\label{eq:ac_nom}
\overline{{\bf A}}_{_{{\bf R}_{k+1}}} = \frac{{\bf A}_{{\bf R}_{k+1}}}{{\bf A}_{{\bf R}_{k+1}}(0,0,0)} = \frac{{\bf R}_{k+1} * {\bf R}_{k+1}}{\|{\bf R}_{k+1}\|_F^2}
\end{equation}
where $\|\cdot\|_F$ denotes the matrix Frobenius norm. All entries $\overline{{\bf A}}_{\bf rr}(n,p,q)$ satisfies:
\begin{equation}\label{eq:ac_pro_nom}
\lim_{L\to\infty} \overline{{\bf A}}_{_{{\bf R}_{k+1}}}(n,p,q) \approx  \begin{cases}
\ 1 \ &\text{if}\ (n,p,q) = (0,0,0)\\
\ 0 \ &\text{if}\ (n,p,q) \ne (0,0,0)
\end{cases}
\end{equation}
We can now introduce the $\sigma$-independent non-negative scalar measure of 3D residual whiteness defined as:
\begin{equation}\label{eq:ac}
{\mathcal{W}}({\bf R}_{k+1}) = ||\overline{{\bf A}}_{_{{\bf R}_{k+1}}}||_F^2 = \frac{||{\bf R}_{k+1} * {\bf R}_{k+1}||_F^2}{||{\bf R}_{k+1}||_F^4}
\end{equation}

% \vspace{0.5mm}
\subsubsection{Penalty parameter estimation}
Solution ${\bf x}_{k+1}$ of~\eqref{eq:x_solution} actually depends on parameter $\rho_k$ setting. To devise the parameter selection procedure, we make $\rho_k$ explicit by writing $\mathbf{x}_{k+1, \rho_k}$. In order to automatically estimate the penalty parameter $\rho_k$ in~\eqref{eq:stepa}, the term $\|{\bf x}-{\tilde{\bf x}}_{k}\|^{2}$ can be viewed as a regularizer that enforces the solution $\mathbf{x}_{k+1, \rho_k}$ to tend to ${\tilde{ \mathbf{x}}}_{k}$. 
As the restored image ${\bf x}_{k+1, \rho_k}$ tends to fit the desired target image, the related residual image ${\mathbf{r}}_{k+1, \rho_k} = {\bf H}{\mathbf{x}}_{k+1, \rho_k} - {\bf y}$ tends to be close to the Gaussian noise perturbation $\bf n$ in~\eqref{eq:3d}.
With~\eqref{eq:ac}, we propose to estimate the optimal penalty parameter by solving the following scalar optimization problem:
\begin{equation}\label{eq:rho}
\rho_k^{*} = \arg\mathop{\min}_{{\rho_k}}\,{\mathcal{W}}({\mathbf{r}}_{k+1, \rho_k})
\end{equation}
The varying range of $\rho_k$ is $(0, \infty)$. In practice, we substitute the $\infty$ by a sufficiently large value.

% {\bf Theorem 2.} If ..., \textit{${\mathcal{W}}({\mathbf{r}}_{k+1, \rho_k})$ admits a minimum.}

% \textit{Proof.} The proof is declared in detail in Appendix~\ref{appendix:b}.

A fast golden-section search method is used for determining a local minimum of~\eqref{eq:rho}. This method operates iteratively over an interval $(a, b)$ and generates two internal points:
\begin{equation}\label{eq:golden}
\begin{split}
&\rho_k^{(1)} = a + \delta(b - a)\\
&\rho_k^{(2)} = b - \delta(b - a)\\
\end{split}
\end{equation}
where  $\delta =0.618$ is the golden ratio. As shown in
Algorithm~\ref{alg_1}, whiteness criterion $\mathcal{W}({\mathbf{r}}_{k+1, \rho_k})$ is compared at $\rho_k^{(1)}$ and $\rho_k^{(2)}$. If it is smaller at the former point than at the latter point, then $b$ is substituted by $\rho_k^{(2)}$. Otherwise, $a$ is substituted by $\rho_k^{(1)}$. This procedure is repeated with the new smaller interval $(a, b)$ until $b - a < \epsilon$ with $\epsilon$ a small positive threshold. Finally, the estimated optimal penalty parameter is given by:
\begin{equation}\label{eq:est_opt}
\rho_k^{*} = (a+b) / 2
\end{equation}
%The evaluated values of ${\mathcal{W}}({\mathbf{r}}_{k+1, \rho_k^{(1)}})$ and ${\mathcal{W}}({\mathbf{r}}_{k+1, \rho_k^{(2)}})$ are then compared and, if ${\mathcal{W}}({\mathbf{r}}_{k+1, \rho_k^{(1)}}) < {\mathcal{W}}({\mathbf{r}}_{k+1, \rho_k^{(2)}})$, then $\rho_k^{(2)}$ replaces $b$ (else, $\rho_k^{(1)}$ replaces $a$). This procedure is repeated in the new smaller interval $(a, b)$ until $b - a < \epsilon$ with $\epsilon$ a small positive threshold. Finally, the estimated optimal penalty parameter is given by
and the solution of sub-problem~\eqref{eq:stepa} is provided by:
\begin{equation}\label{eq:est_opt_solution}
{\bf x}_{k+1} = ({\bf H}^{T}{\bf H}+\rho_k^* {\bf I})^{-1}({\bf H}^{T}{\bf y}+\rho_k^*{\bf \tilde{\bf x}}_{k})
\end{equation}
\renewcommand{\algorithmicrequire}{ \textbf{Input:}} %Use Input in the format of Algorithm
\renewcommand{\algorithmicensure}{ \textbf{Output:}} %UseOutput in the format of Algorithm
\begin{algorithm}[!t]
	\caption{Adaptive Penalty Parameter Estimation.}
	\label{alg_1}
	\begin{algorithmic}
		\REQUIRE Blurred observation $\bf y$, internal image ${\tilde{\bf x}}_{k}$,\\ blurring kernel $\bf H$.\\
		\ENSURE Optimal adaptive parameter $\rho_k^*$.\\
		\STATE Initialize $a, b , \epsilon$.
		\WHILE{$b-a>\epsilon$}
		\vspace{1mm}
		\STATE $\rho_k^{(1)} = a + \delta(b - a)$
		\vspace{1mm}
		\STATE $\rho_k^{(2)} = b - \delta(b - a)$
		\STATE \bf{if} ${\mathcal{W}}({\mathbf{r}}_{k+1, \rho_k^{(1)}}) < {\mathcal{W}}({\mathbf{r}}_{k+1, \rho_k^{(2)}})$
				\vspace{1mm}
		\STATE $\qquad b = \rho_k^{(2)}$
		\STATE \bf{else}
		\STATE $\qquad a = \rho_k^{(1)}$
		\ENDWHILE
	\STATE $\rho_k^{*} = (a+b) / 2$
% 			\vspace{1mm}
% 	\STATE ${\bf x}_{k+1} = ({\bf H}^{T}{\bf H}+\rho_k^* {\bf I})^{-1}({\bf H}^{T}{\bf y}+\rho_k^*{\bf \tilde{\bf x}}_{k})$
	\end{algorithmic}
\end{algorithm}
% \vspace{0.5mm}

\subsubsection{Stopping criterion} 
To take both HSI deconvolution performance and computational time into account, it is important to properly set the maximum number of iterations. Iterations can be performed until no significant improvement between two consecutive iterations is observed. Considering the whiteness measure in~\eqref{eq:ac}, we propose to stop the iterative process with the following normalized criterion:
\begin{equation}\label{eq:stop}
{\mathcal{W}}({\mathbf{r}}_{k+1}) \ge {\mathcal{W}}({\mathbf{r}}_{k}) \text{\quad or\quad} \frac{||{\mathcal{W}}({\mathbf{r}}_{k+1})  -  {\mathcal{W}}({\mathbf{r}}_{k})||}{{\mathcal{W}}({\mathbf{r}}_{k+1})} < \zeta
\end{equation}
where $\zeta$ is a small positive threshold, ${\mathbf{r}}_{k}$ and ${\mathbf{r}}_{k+1}$ represent the residual image of the solutions ${\mathbf{x}}_{k}$ and ${\mathbf{x}}_{k+1}$, respectively. 

\subsection{Learning spectral-spatial priors via B3DDN}
\label{ssec:subhead}

Instead of using an handcrafted regularizer $\Phi(\cdot)$ and solving subproblem~\eqref{eq:stepb} explicitly, we propose to carry out this task with a deep neural network based denoiser. This denoiser is trained beforehand to extract spectral-spatial prior information from hyperspectral training observations. Then it is plugged into the iterative algorithm to solve subproblem~\eqref{eq:stepb}. We denote this denoising operator by $\mathcal{D}(\cdot)$. As it is performed in the 3D image domain to jointly capture spatial and spectral information, we write~\eqref{eq:denoise} as follows:
\begin{equation}\label{eq:denoiser}
{\bf Z}_{k+1} = \mathcal{D}({\bf \tilde{Z}}_k, \sigma_k)
\end{equation}
Observe that $\mathcal{D}(\cdot)$ is parameterized by the noise level $\sigma_k$. For setting it, most existing methods use empirical strategies that may lead to under-denoising or over-smoothing of ${\bf \tilde{Z}}_k$~\cite{chen2020learning}. In addition, since $\sigma_k$ decreases as iterations progress, some works choose to train a set of specific models that can handle different noise levels~\cite{zhang2017learning}. To avoid these redundant learning tasks, we shall now see how to design a blind 3D denoising network $\mathcal{F}(\cdot)$ with respect to $\sigma_k$, but parameterised by $\Theta$, by considering residual learning formulation:
\begin{equation}
    \label{eq:denoiser_blind}
    {\bf Z}_{k+1} = {\bf \tilde{Z}}_k 
    - \mathcal{F}({\bf \tilde{Z}}_k;\Theta)
\end{equation}

%\begin{figure}[]
%	\centering
%	\includegraphics[scale=0.8]{Figures/3D.pdf}
%	\caption{Difference between 2D and 3D convolution.}
%	\label{fig:2D3D}
%\end{figure}
\subsubsection{3D convolution}
Unlike 2D convolution resulting in spectral information distortion, 3D convolution extracts spatial features from neighboring pixels and spectral features from adjacent bands, simultaneously, without compromising spectral resolution. 3D convolution also involves less parameters, and it is more appropriate for hyperspectral image processing due to the difficulty in capturing a big enough volume of hyperspectral data.
In addition, 3D convolution enables the neural network to handle HSIs with arbitrary number of spectral bands without modifying its architecture~\cite{liu20193}. In this way, there is no need to retrain a neural network when the number of spectral bands changes. This key property allows our method to be applied to any real-world dataset by using a pre-trained neural network.

\subsubsection{Network architecture}
The B3DDN architecture is illustrated in Fig.~\ref{fig:net} (top). Each 3D-block contains a 3D convolution layer (3DConv), a batch normalization (BN) layer and a ReLU layer. Batch normalization is used to speed up the training process as well as to boost the denoising performance~\cite{zhang2017beyond}. Besides the input layer and the output layer, a 3D convolution layer (3DConv), a ReLU activation function layer,  $B$  3D-blocks and a last 3D convolution layer are sequentially connected to form the proposed network. The last convolutional layer contains one 3D-filter while the others are composed of 32 3D-filters. The kernel size of each 3D-filter is 3$\times$3$\times$3, which means that the depth of the kernel along the spectral dimension and its size over the spatial dimension are 3 and 3$\times$3 respectively. Compared to existing complex network architectures for HSI denoising, B3DDN achieves satisfactory performance with less parameters. Moreover, it enables us to apply the neural network learned with simulated data, to real data that lacks ground truth. An example is provided in subsection~\ref{subsec:real}.

\subsubsection{Learning strategy}
The input of the proposed B3DDN is a noisy hyperspectral image ${\bf \tilde{z}} = {\bf{z}} + {\bf{v}}$, where $\bf v$ is a Gaussian noise with arbitrary standard deviation. Inspired by 2D image denoising algorithm~\cite{zhang2017beyond}, we consider the learning residual to predict the residual error $\mathcal{F}({\bf \tilde{z}}_k;\Theta) \approx \bf v$ in our denoising network. Then we can achieve the estimated clean image by ${\bf \tilde{z}} - \mathcal{F}({\bf \tilde{z}};\Theta)$. To train the blind neural network $\mathcal{F}(\ \cdot\ ;\Theta)$, we use the following loss function:
\begin{equation}\label{eq:loss}
\ell(\Theta) = \|\mathcal{F}({\bf \tilde{z}}_m;\Theta)-({\bf \tilde{z}}_m-{\bf z}_m)\|_{1}
\end{equation}
where $\{({\bf \tilde{z}}_m, {\bf z}_m)\}_{m=1}^{M}$ is a training set of generated noisy-clean HSI (patch) pairs with various noise levels. 
%Note that the $\ell_1$-norm is used as we found to lead to better performance than the $\ell_2$-norm in hyperspectral image denoising. 
{Note that the $\ell_1$-norm is used as a loss that is more robust to noise than the $\ell_2$-norm, found providing better performance in image restoration in the literature~\cite{zhao2016loss, wang2021hyperspectral}.}
After the B3DDN has been trained, it is incorporated into the ADMM framework as a blind denoiser, yielding Algorithm~\ref{alg_2}.

\renewcommand{\algorithmicrequire}{ \textbf{Input:}} %Use Input in the format of Algorithm
\renewcommand{\algorithmicensure}{ \textbf{Output:}} %UseOutput in the format of Algorithm
\begin{algorithm}[!t]
	\caption{Tuning-free HSI deconvolution with deep priors learnt from B3DDN.}
	\label{alg_2}
	\begin{algorithmic}
		\REQUIRE Network parameters $\Theta$, blurred observation $\bf y$,\\ blurring kernel $\bf H$.\\
		\ENSURE Deblurred HSI $\bf X$.\\
		\STATE Initialize $\bf x=x_0$, auxiliary variable $\bf z_0=x_0$,\\ scaled dual variable $\bf u_0=0$, $k=0$.
		\WHILE{Stopping criteria in~\eqref{eq:stop} are not met}
		\STATE ${\tilde{\bf x}}_{k} = {\bf z}_k - {\bf u}_k$
		\STATE Estimate $\rho_k^* $ using Algorithm~\ref{alg_1}
		\vspace{1mm}
		\STATE ${\bf x}_{k+1}=({\bf H}^{T}{\bf H}+\mu {\bf I})^{-1}({\bf H}^{T}{{\bf y}}+\rho_k^* {\bf \tilde{\bf x}}_{k})$
		\STATE ${\tilde{\bf z}}_{k} = {\bf x}_{k+1} + {\bf u}_k$
		\vspace{1mm}
		\STATE ${\bf Z}_{k+1} = {\bf \tilde{Z}}_k - \mathcal{F}({\bf \tilde{Z}}_k;\Theta)$
		\STATE ${\bf u}_{k+1} = {\bf u}_k + {\bf x}_{k+1} - {\bf z}_{k+1}$
		\STATE $k = k + 1$
		\ENDWHILE
	\end{algorithmic}
\end{algorithm}

\begin{figure*}[!t]
	\centering
	\includegraphics[trim = 0mm 0mm 0mm 0mm, clip, scale=0.46]{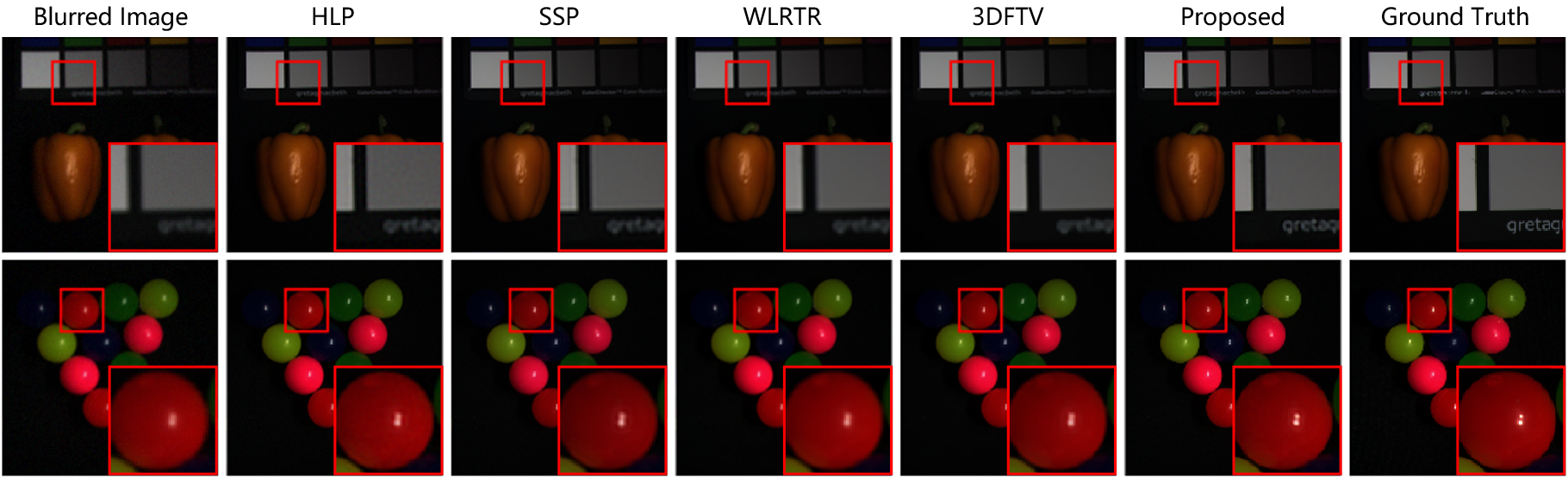}
	\caption{{Visual results for all methods in the blurring scenario (a) on the CAVE dataset.  The first and second rows present the results for two different blurred images. The false color images were generated for clear visualization with the 22nd, 14th and 7th channels used for red,  green and blue, respectively.}}
	\label{fig:cave}
\end{figure*}

\begin{figure}[!t]
	\centering
	\includegraphics[trim = 0mm 1mm 0mm 1mm, clip, scale=0.72]{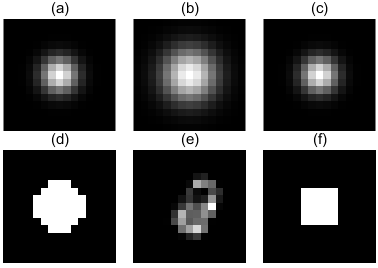}
	\caption{Blurring kernels used for the experiments: (a)-(c) are Gaussian kernels, (d)-(f) are circle, motion and square kernels respectively. }
	\label{fig:kernel}
\end{figure}

\section{Experiments}
\label{sec:experiment}
In this section, we shall conduct experiments of HSI deconvolution on both simulated and real-world datasets to validate our method. The results provided by the proposed method are compared with those of several HSI deconvolution methods from both quantitative
and qualitative perspectives. The source code and the proposed real-world data is made available at \url{https://github.com/xiuheng-wang/Tuning_free_PnP_HSI_deconvolution}.

\subsection{Simulation datasets and experimental setup}
Two simulation datasets, on the one hand the Columbia Multispectral Database (CAVE)\footnote{https://www1.cs.columbia.edu/CAVE/databases/multispectral/}~\cite{yasuma2010generalized}, and on the other hand a remotely sensed hyperspectral data over Chikusei\footnote{http://naotoyokoya.com/Download.html}~\cite{NYokoya2016}, were used to evaluate the performance of our method.

\subsubsection{CAVE dataset}
The CAVE dataset contains 32 HSIs recorded under controlled illuminations in a laboratory. Each image has a spatial resolution of $512\times 512$ pixels, over 31 spectral channels ranging from 400 nm to 700 nm at a wavelength interval of 10 nm. 
\subsubsection{Chikusei dataset}
The Chikusei dataset is an airborne hyperspectral scene acquired by a Visible and Near-Infrared imaging sensor over agricultural and urban regions in Chikusei,
Ibaraki, Japan. The scene consists of $2517\times 2335$ pixels with a ground sampling distance of 2.5 m, over 128 spectral channels ranging from 363 nm to 1018 nm. The black boundaries in the spatial domain were removed, leading to a scene of size $2048\times 2048$ pixels.

\begin{table*}[]
	\centering
	\caption{RMSE, PSNR, SSIM and ERGAS of the different methods applied to the CAVE dataset in the 6 blurring scenarios.}
	{\begin{tabular}{c|c|ccccc}
		\hline \hline
		{Scenarios} & {Metrics}                                           
		& HLP & SSP & WLRTR & 3DFTV & Ours   \\ \hline \hline
		\multirow{4}{*}{(a)}   
		& RMSE & 4.420 $\pm$ 1.787 & 4.848 $\pm$ 1.825 & 4.735 $\pm$ 2.076 & 4.332 $\pm$ 1.863 & \textbf{3.132 $\pm$ 1.320} \\      
		& PSNR &36.166  $\pm$   {\bf3.334} & 35.373  $\pm$  3.385&35.872  $\pm$  3.759 & 36.450  $\pm$  3.793 & {\bf 39.252}  $\pm$  3.465\\
 		%& SAM  &11.32  $\pm$  6.32  & 9.78  $\pm$  4.73&\textbf{6.05  $\pm$  2.83} &  8.16  $\pm$  5.67\\
		& SSIM &0.9167  $\pm$  0.0379  & 0.9305  $\pm$  0.0393 & 0.9380  $\pm$  0.0466 & 0.9401 $\pm$ 0.0439  & {\bf 0.9493  $\pm$  0.0367}\\
		& ERGAS  & 18.15  $\pm$  8.25  & 19.51  $\pm$  8.12&18.96  $\pm$  8.33  & 17.34 $\pm$ 7.69 & {\bf 13.01   $\pm$  6.23}\\
		\hline
		\multirow{4}{*}{(b)}     
		& RMSE & 5.707 $\pm$ 2.452 & 5.955 $\pm$ 2.398 & 6.439 $\pm$ 2.812 & 5.667  $\pm$ 2.539 & \textbf{4.581 $\pm$ 1.993} \\     
		& PSNR &34.034  $\pm$  3.567  & 33.541  $\pm$  {\bf3.492} & 33.084  $\pm$  3.740& 34.116  $\pm$  3.872 & {\bf36.305}  $\pm$  3.612\\
 		%& SAM  &11.30  $\pm$  6.02  & 10.26  $\pm$  4.66 &{\bf6.63  $\pm$  2.78} & 8.55  $\pm$  5.18 \\
		& SSIM &0.8911  $\pm$  0.0483  & 0.9031  $\pm$  0.0494 &0.9025  $\pm$  0.0616 & 0.9136 $\pm$ 0.550 & {\bf0.9234  $\pm$  0.0422 }\\
		& ERGAS  & 22.92  $\pm$  10.11 &23.71  $\pm$  9.86 & 25.46  $\pm$  10.84 & 22.40 $\pm$ 9.86 & {\bf18.54  $\pm$  8.39}\\
		\hline
		\multirow{4}{*}{(c)} 
		& RMSE & 7.669 $\pm$ 1.390 & 5.270 $\pm$ 1.622 & 5.099 $\pm$ 1.972 & 5.016 $\pm$ 1.727  &\textbf{4.225 $\pm$ 1.324} \\     
		& PSNR &30.599  $\pm$ {\bf 1.550} & 34.309  $\pm$  2.607 & 34.827 $\pm$ 3.201 &34.741 $\pm$ 2.975 & {\bf36.211}  $\pm$ 2.485 \\
 		%& SAM  &24.68  $\pm$  9.32  & 14.01 $\pm$ 6.81 & {\bf 11.57 $\pm$ 5.99}  &12.96  $\pm$ 6.91 \\
		& SSIM &0.6406  $\pm$  0.0337  & 0.8565 $\pm$  0.0539&{\bf 0.8956 $\pm$ 0.0387} & 0.8851  $\pm$ 0.0390 & 0.8708  $\pm$ 0.0594 \\
		& ERGAS  & 33.49  $\pm$  16.27 & 22.28 $\pm$  10.14& 20.80 $\pm$ \textbf{9.07} & 20.47 $\pm$ 8.66 &{\bf18.64}  $\pm$  9.28\\
		\hline
		\multirow{4}{*}{(d)} 
		& RMSE & 4.189 $\pm$ 1.636 & 4.584 $\pm$ 1.680 & 4.328 $\pm$ 1.903 & 4.167 $\pm$ 1.803  &\textbf{2.305 $\pm$ 0.938} \\     
		& PSNR & 36.548  $\pm$  3.181 & 35.862 $\pm$ 3.331 & 36.686 $\pm$ 3.736  & 36.805 $\pm$ 3.803 & {\bf41.653} $\pm$ \textbf{3.074} \\
 		%& SAM  &11.88  $\pm$  6.58  & 9.71 $\pm$ 4.75 & {\bf 6.14 $\pm$ 3.12} & 9.23 $\pm$ 7.25 \\
		& SSIM &0.9165  $\pm$  0.0348 & 0.9354 $\pm$ 0.0374 & 0.9450 $\pm$ 0.0436 & 0.9403 $\pm$ 0.0430 &{\bf 0.9542 $\pm$ 0.0340} \\
		& ERGAS  & 17.36  $\pm$  7.98 & 18.49 $\pm$  7.67& 17.45 $\pm$ 7.82 & 16.69 $\pm$ 7.46 &{\bf9.86 $\pm$ 5.17} \\
		\hline
		\multirow{4}{*}{(e)}   
		& RMSE & 3.759 $\pm$ 1.166 & 3.954 $\pm$ \textbf{1.333} & 4.335 $\pm$ 1.780 & 3.587 $\pm$ 1.443 & \textbf{3.041} $\pm$ 2.783 \\       
		& PSNR &37.149  $\pm$  {\bf2.492} &37.160  $\pm$  3.108& 36.497 $\pm$ 3.490 & 37.991 $\pm$ 3.543 & {\bf40.722} $\pm$ 5.730 \\
		& SSIM &0.9118  $\pm$  \textbf{0.0239} & \textbf{0.9472} $\pm$ 0.0311 & 0.9428 $\pm$ 0.0436 & 0.9510 $\pm$ 0.0397 & {0.8907 $\pm$ 0.1642} \\
		& ERGAS  & 15.94  $\pm$  7.39 & 16.01 $\pm$ \textbf{6.56} & 17.46 $\pm$ 7.49 & 14.37 $\pm$ 6.16 & {\bf15.56} $\pm$ {19.66} \\
		\hline
		\multirow{4}{*}{(f)} 
		& RMSE & 3.971 $\pm$ 1.453 & 4.356 $\pm$ 1.563 & 4.109 $\pm$ 1.765  & 3.957 $\pm$ 1.666 & \textbf{2.280 $\pm$ 1.231} \\      
		& PSNR &36.910  $\pm$ {\bf 2.985 } & 36.322 $\pm$  3.302& 37.130 $\pm$ 3.698 & 37.225 $\pm$ 3.743 & {\bf41.932} $\pm$ 3.687 \\
		& SSIM &0.9195  $\pm$  0.0270 &0.9397  $\pm$ 0.334 & 0.9480 $\pm$ 0.0450 & 0.9468 $\pm$ 0410 & {\bf0.9475 $\pm$ 0.0618} \\
		& ERGAS  & 16.58  $\pm$  7.61 &17.60  $\pm$ 7.26 & 16.64 $\pm$ 7.46 & 15.89 $\pm$ 7.04 &{\bf9.79 $\pm$ 5.89} \\
		\hline \hline
	\end{tabular}}
	\vspace{1mm}
	\\
	{The best results are indicated by boldface numbers.}
	\label{tab_1}	
\end{table*}

\begin{table}[]
	\centering
	{\caption{RMSE, PSNR, SSIM and ERGAS of the different methods applied to the CAVE dataset in the blurring scenario (a) with various noise levels.}
	\begin{tabular}{c|c|cccc}
		\hline \hline
		{$\sigma$} & {Methods}                                           
		& RMSE & PSNR & SSIM & ERGAS  \\ \hline \hline
		\multirow{5}{*}{0.01}   
		& HLP & 4.420  &36.166   & 0.9167 & 18.15 \\      
		& SSP & 4.848   & 35.373   & 0.9305 & 19.51 \\
		& WLRTR & 4.735  & 35.872   & 0.9380   &18.96\\
		& 3DFTV  & 4.332  & 36.450     & 0.9401    & 17.34 \\
		& Ours & \textbf{3.132}  & {\bf39.252}  & {\bf0.9493}   & {\bf13.01} \\
		\hline
		\multirow{5}{*}{0.02}   
		& HLP & 5.597  & 33.571  & 0.8084 & 23.77 \\      
		& SSP &  5.001   & 34.951   & 0.8973 & 20.52 \\
		& WLRTR & 4.817  & 35.602  & 0.9283   &19.39\\
		& 3DFTV  & 4.486  & 36.006    & \textbf{0.9301}   & 17.96 \\
		& Ours & \textbf{3.574}  & {\bf37.851}   & {0.9140}   & {\bf15.17} \\
		\hline
		\multirow{5}{*}{0.03}   
		& HLP & 7.669  &30.599  & 0.6406 & 33.49 \\      
		& SSP  & 5.270   & 34.309   & 0.8565 & 22.28\\
		& WLRTR & 5.099  & 34.827   &{\bf0.8956}   & 20.80\\
		& 3DFTV  & 5.016  &34.741    & 0.8851    & 20.47 \\
		& Ours & \textbf{4.225}  & {\bf36.211}   & 0.8708  &{\bf18.64} \\
		\hline
		\multirow{5}{*}{0.04}   
		& HLP & 10.018  &28.206  & 0.4942 & 44.43 \\      
		& SSP  & 5.643   & 33.547   & 0.8155 & 24.66 \\
		& WLRTR & 6.611  & 32.107   & 0.7539   & 27.42 \\
		& 3DFTV  & 6.143  & 32.682  & 0.7778 & 26.05 \\
		& Ours & \textbf{4.750}  &  {\bf35.060}   & \textbf{0.8324}  & {\bf21.37}\\
		\hline
		\multirow{5}{*}{0.05}   
		& HLP & 12.411  &26.320  & 0.3859  &  55.50\\      
		& SSP  & 6.101   & 32.742  & 0.7766 & 27.49 \\
		& WLRTR & 9.496   &  28.748  & 0.5363  &  40.74 \\
		& 3DFTV  & 7.696  &  30.572   & 0.6462 & 33.57   \\
		& Ours & \textbf{5.329}  & {\bf33.976}   & \textbf{0.7984}  & {\bf24.60}\\
		 \hline
		 \hline
	\end{tabular}}
	\vspace{1mm}
	\\
	{The best results are indicated by boldface numbers.}
	\label{tab_noise}	
\end{table}

The HSIs of the two datasets were scaled to the range $[0, 1]$, and then used as ground truths for $\bf x$. The observations $\bf y$ were generated by using the blurring kernels $\bf H$ and corrupted with a white Gaussian noise $\bf n$ with standard deviation $\sigma$, with $\bf H$ and $\sigma$ defined as follows; see Fig.~\ref{fig:kernel}:
\begin{itemize}
	\item[(a)] 9$\times$9 Gaussian kernel with bandwidth 2, and $\sigma\!=\!0.01$;
	\item[(b)] 13$\times$13 Gaussian kernel with bandwidth 3, and $\sigma \!=\!0.01$;
	\item[(c)] 9$\times$9 Gaussian kernel with bandwidth 2, and $\sigma \!=\! 0.03$;
	\item[(d)] Circle kernel with diameter 7, and $\sigma = 0.01$;
	\item[(e)] Motion kernel from \cite{levin2009understanding} of size 13$\times$13, and $\sigma =$ 0.01;
	\item[(f)] Square kernel with side length 5, and $\sigma =$ 0.01.
\end{itemize}
The first 20 images were selected
from the CAVE dataset for training and the remaining 12 images were used for the
test. For the Chikusei dataset, a $1024\times 2048$ sub-image was extracted from the top area of the image for training while the remaining part was cropped into 32 non-overlapping $256\times 256\times128$ sub-images
that were used as test data.

\subsection{Implementation details}
We implemented the proposed blind denoising network B3DDN
with PyTorch framework. The Adam optimizer~\cite{kingma2014adam} with an initial learning rate 0.0002 and batch size 64 was used to minimize the loss function~\eqref{eq:loss} with 500 epochs. The weights were initialized by the method in~\cite{he2015delving}. At every epoch of the training stage, each original HSI was randomly cropped into 128 and 512 patches of size 64$\times$64 respectively for the CAVE and the Chikusei datasets. To train the B3DDN in a blind manner, we added an i.i.d. Gaussian noise with random standard deviation in the range $[0.2, 10]$ to each patch, which was then randomly rotated or flipped for data augmentation purpose. 
We set the number $B$ of 3D blocks to $8$ by considering the computational cost and memory demand, and thus the number of parameters of the proposed B3DDN denoiser is 10113.

Once the denoiser was trained, assuming that the statistics of the test images differ from training images, we plugged the B3DDN into the ADMM. Since the computational complexity of 3D discrete correlation in~\eqref{eq:ac} can be high ($\mathcal{O}(L^2)$), we used the fast Fourier transform ($\mathcal{O}(L\log{}L)$) to compute it. 
% To further accelerate the calculation, we only compute the correlation of the spatially center $100\times100$ image region. 
Step~\eqref{eq:est_opt_solution} was also efficiently computed in the Fourier domain.
For the golden-section search method and the stopping criterion presented in Subsection~\ref{ssec:subhead_2}, we set $a = 0$, $b = 10$, $\epsilon = 0.001$ and $\zeta = 0.0002$.

\subsection{Quantitative metrics and baselines}
{In order to evaluate the quality
of the deconvolution result $\widehat{\bf X}$ by comparing it with the ground truth
of ${\bf X}$, we considered four quantitative metrics.
The first one is the Root Mean-Square Error (RMSE), dedined as
\begin{align}
	\text{RMSE}
	{}={} \sqrt{\!\frac{1}{NPQ}\!\sum_{i=1}^{N}\big\|{\widehat{\bf X}_i -{\bf X}_i\big\|_F^2}}
    \nonumber\,,
\end{align}
which measures the similarities between the deconvolution image and the reference image. A lower RMSE value indicates better quality. 
The second metric is the Peak-Signal-to-Noise-Ratio (PSNR):
\begin{align}
	\text{PSNR}
	{}={} \frac{1}{N} \sum_{i=1}^{N}
    10\log_{10} \Bigg(
    \frac{PQ\, \max({\bf X}_i)^2}{\big\|{\widehat{\bf X}_i -{\bf X}_i\big\|_F^2}} \Bigg)
    \nonumber\,,
\end{align}
which measures the quality of the decovolution image compared to the original image. The higher the PSNR, the better quality. 
The third metic is the average of Structural SIMilarity (SSIM)~\cite{wang2004image}, averaged over all channels of $\widehat{\bf X}$ and ${\bf X}$, i.e., 
% \begin{align}
% \text{SSIM}
% 	{}={} \!\frac{1}{N} \! \sum_{i=1}^{N}\mathrm{SSIM} (\widehat{\bf X}_i, {\bf X}_i)
%     \nonumber \,.
% \end{align}
\begin{align}
\text{SSIM}
	{}={} \!\frac{1}{N} \! \sum_{i=1}^{N}\frac{(2\mu_{\widehat{\bf X}_i} \mu_{{\bf X}_i} + C_1 )(2\sigma_{\widehat{\bf X}_i{\bf X}_i} + C_2)}{(\mu_{\widehat{\bf X}_i} + \mu_{{\bf X}_i} + C_1)(\sigma_{\widehat{\bf X}_i} + \sigma_{{\bf X}_i} + C_2)}
    \nonumber \,,
\end{align}
where  $\mu_{\widehat{\bf X}_i}$ and $\mu_{{\bf X}_i}$ are the mean values of images $\widehat{\bf X}_i$ and ${\bf X}_i$, $\sigma_{\widehat{\bf X}_i}$ and $\sigma_{{\bf X}_i}$ are the standard deviations of $\widehat{\bf X}_i$ and ${\bf X}_i$, $\sigma_{\widehat{\bf X}_i{\bf X}_i}$ is the covariance
of $\widehat{\bf X}_i$ and ${\bf X}_i$, and $C_1 > 0$ and $C_2 > 0$ are constants. The SSIM is an indicator of the spatial structure preservation of the deconvolution image. A higher the SSIM value indicates better spatial structure preservation.
The last metric is the Erreur Relative Globale Adimensionnelle de Synth\`{e}se (ERGAS)~\cite{wald2000quality} defined as
\begin{align}
\text{ERGAS}
	{}={} 100 \sqrt{\!\frac{1}{N} \! \sum_{i=1}^{N}  \frac{\big\|{\widehat{\bf X}}_i -{\bf X}_i\big\|_F^2}{\mathrm{mean}({\widehat{\bf X}}_i)^2}} 
    \nonumber \,,
\end{align}
which charactersizes the overall quality of the deconvolution image. A smaller ERGAS means a better result. }

We compared our method with three HSI deconvolution methods of reference: {hyper-laplacian priors (HLP)~\cite{krishnan2009fast}, spatial and spectral priors (SSP)~\cite{henrot2012fast}, weighted low-rank tensor recovery (WLRTR)~\cite{chang2020weighted}, 3D fractional total variation (3DFTV)~\cite{guo2021three}}, each with well-designed regularizers. The HLP considers spatial gradient priors, i.e., the hyper-Laplacian priors of images. The SSP exploits both the spatial and spectral smoothness priors of hyperspectral images. The WLRTR simultaneously captures non-local similarity within spectral-spatial cubic and spectral correlation by a low-rank tensor recovery model. {The 3DFTV exploits both the local and non-local smoothness of images in all dimensions.} We used the codes provided by the authors of these methods and downloaded them, and we tuned their parameters by following the rules as stated in the corresponding papers to achieve the best deconvolution performance. 

\begin{figure*}[!t]
	\centering
	\includegraphics[trim = 0mm 0mm 0mm 0mm, clip, scale=0.46]{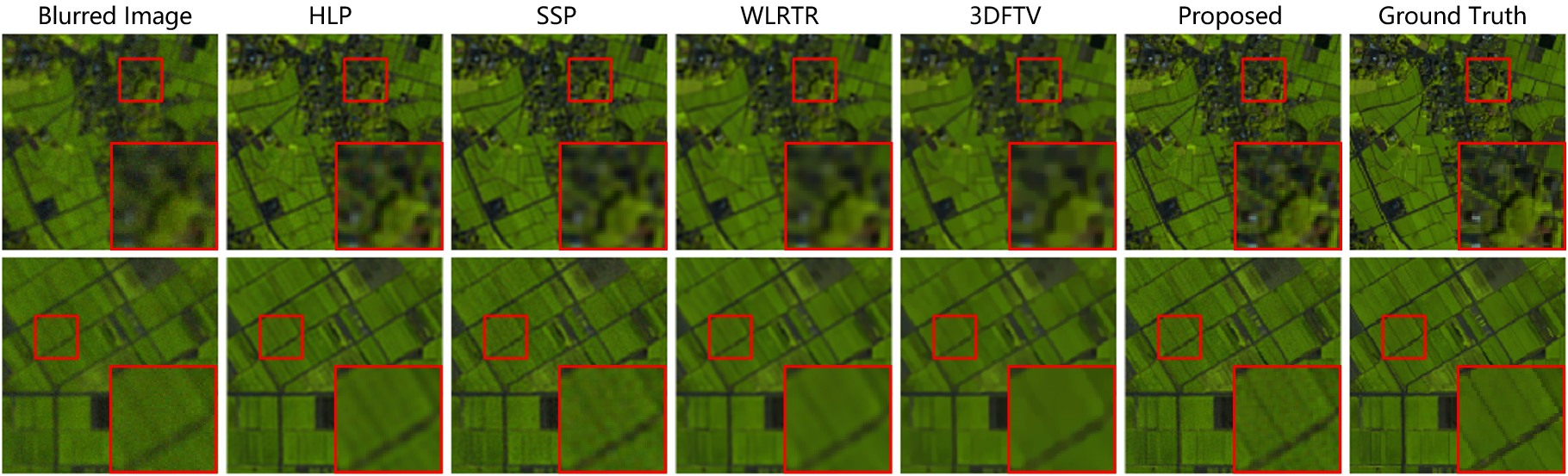}
	\caption{{Visual results for all methods with the blurring scenario (d) applied to the Chikusei dataset. The first and second rows present the results for two different images. The false color images were generated for clear visualization with the 122nd, 84th and 57th channels used for red, green and blue, respectively.}}
	\label{fig:chikusei}
\end{figure*}

\begin{table*}[!t]
	\centering
	\caption{RMSE, PSNR, SSIM and ERGAS of the different methods applied to the Chikusei dataset in the 6 blurring scenarios.}
	{\begin{tabular}{c|c|ccccc}
		\hline \hline
		{Scenarios} & {Metrics}                                           
		& HLP & SSP & WLRTR & 3DFTV & Ours   \\ \hline \hline
		\multirow{4}{*}{(a)}  
		& RMSE & 3.233 $\pm$ 0.420 & 3.050 $\pm$ 0.452 & 3.138 $\pm$ 0.518 & 3.207 $\pm$ 0.501 & \textbf{2.560 $\pm$ 0.316} \\     
		& PSNR &38.979  $\pm$  1.131 & 40.182 $\pm$ 1.528 & 40.051 $\pm$ 1.710 & 39.546 $\pm$ 1.583 & \textbf{41.032} $\pm$ \textbf{1.076}  \\
% 		& SAM  &4.65  $\pm$  0.59 & 3.18 $\pm$ 0.50 & 3.11 $\pm$ 0.55 & \textbf{3.93 $\pm$  0.49} \\
		& SSIM &0.9124  $\pm$  0.0141 & 0.9334 $\pm$ 0.0148 & 0.9267 $\pm$ 0.0183 & 0.9171 $\pm$ 0.0191 &\textbf{0.9420  $\pm$  0.0086}\\
		& ERGAS  & 32.25  $\pm$  3.40 & 28.13 $\pm$ \textbf{2.64} & \textbf{25.29} $\pm$ 3.54 & 35.37 $\pm$ 2.93 & 27.87  $\pm$  2.84 \\
		\hline
		\multirow{4}{*}{(b)} 
		& RMSE & 3.945 $\pm$ 0.576 & 3.819 $\pm$ 0.566 & 4.091 $\pm$ 0.678 & 4.037 $\pm$ 0.644 & \textbf{3.428 $\pm$ 0.502} \\      
		& PSNR &37.604  $\pm$  \textbf{1.348} & 38.392 $\pm$ 1.573 & 37.872 $\pm$ 1.741 & 37.708 $\pm$ 1.637& \textbf{38.989} $\pm$ 1.438 \\
		% & SAM  &4.89  $\pm$  0.69 & 3.71 $\pm$ 0.65 & 3.84 $\pm$ 0.74 & \textbf{4.14 $\pm$ 0.57} \\
		& SSIM &0.8822  $\pm$  0.0227 &0.9016  $\pm$ 0.0222 & 0.8871 $\pm$ 0.0276 & 0.8819 $\pm$ 0.0275 & \textbf{0.9091 $\pm$ 0.0183} \\
		& ERGAS  & 35.30  $\pm$  4.08 & 32.40 $\pm$ 3.60 & 31.45 $\pm$ 4.97 & 39.85 $\pm$ 3.59 & \textbf{30.92} $\pm$ \textbf{3.23} \\
		\hline
		\multirow{4}{*}{(c)}  
		& RMSE & 7.094 $\pm$ \textbf{0.197} & 3.506 $\pm$ 0.386 & 3.777 $\pm$ 0.429 & 3.662 $\pm$ 0.416 & \textbf{3.413} $\pm$ 0.295 \\     
		& PSNR & 31.391  $\pm$  \textbf{0.252} & \textbf{37.942} $\pm$  0.930 & 37.447 $\pm$ 0.955 & 37.756 $\pm$ 0.994 & 37.934 $\pm$ 0.703 \\
		& SSIM &0.6268  $\pm$  \textbf{0.0060} & 0.8839 $\pm$ 0.0146 & 0.8816 $\pm$ 0.0157 & \textbf{0.8841} $\pm$ 0.0152 & 0.8783 $\pm$ 0.0113 \\
		& ERGAS  & 90.14  $\pm$  11.92 &50.26  $\pm$ 6.80 & \textbf{39.95 $\pm$ 4.34} & 48.15 $\pm$ 5.01 &  51.38 $\pm$ 7.55 \\
		\hline
		\multirow{4}{*}{(d)}  
		& RMSE & 3.361 $\pm$ \textbf{0.177} & 2.879 $\pm$ 0.427 & 2.890 $\pm$ 0.477 & 3.076 $\pm$ 0.483 & \textbf{2.335} $\pm$ 0.202 \\     
		& PSNR &39.122  $\pm$  0.995 & 40.625 $\pm$ 1.505 & 40.724 $\pm$ 1.689 & 39.900 $\pm$ 1.587 & \textbf{41.290 $\pm$ 0.729} \\
		& SSIM &0.9148  $\pm$  0.0114 & 0.9399 $\pm$ 0.0132 & 0.9364 $\pm$ 0.0160 &  0.9228 $\pm$ 0.0178& \textbf{0.9430} $\pm$ \textbf{0.0045} \\
		& ERGAS  & 32.76  $\pm$  3.47 & 27.22 $\pm$ \textbf{2.47} & \textbf{23.73} $\pm$ 3.17 & 34.59 $\pm$ 2.86 & 32.56 $\pm$ 3.76\\
		\hline
		\multirow{4}{*}{(e)}   
		& RMSE  & 2.960 $\pm$ 0.253 & 2.436 $\pm$ 0.352 & 2.790 $\pm$ 0.460 &2.797 $\pm$ 0.436  & \textbf{1.995 $\pm$ 0.106} \\       
		& PSNR &39.127  $\pm$  0.681 &  41.869 $\pm$ 1.380  & 41.025 $\pm$ 1.644 & 40.574 $\pm$ 0.534 & \textbf{42.207 $\pm$ 0.468} \\
		& SSIM &0.9147  $\pm$  0.0055 & \textbf{0.9558} $\pm$ 0.0090  & 0.9408 $\pm$ 0.0147 & 0.9338 $\pm$ 0.0149 & 0.9507 $\pm$ \textbf{0.0028} \\
		& ERGAS  & 35.79  $\pm$  4.15 &   25.42 $\pm$ \textbf{2.14}  & \textbf{23.09} $\pm$ 2.90 & 33.56 $\pm$ 2.96 & 36.06 $\pm$ 4.93 \\
		\hline
		\multirow{4}{*}{(f)}  
		& RMSE & 2.990 $\pm$ 0.327 & 2.688 $\pm$ 0.395 & 2.691 $\pm$ 0.441 & 2.913 $\pm$ 0.453 & \textbf{2.148 $\pm$ 0.186} \\        
		& PSNR &39.352  $\pm$  0.902 & 41.174 $\pm$ 1.473 & 41.313 $\pm$ 1.659 & 40.334 $\pm$ 1.561 & \textbf{41.971} $\pm$ \textbf{0.694} \\
		& SSIM &0.9188  $\pm$  0.0093 & 0.9456 $\pm$ 0.0116 & 0.9438 $\pm$ 0.0140 & 0.9295 $\pm$ 0.0161 & \textbf{0.9506 $\pm$ 0.0038} \\
		& ERGAS  & 32.68  $\pm$  3.53 & 26.19  $\pm$  \textbf{2.29} & \textbf{22.46} $\pm$ 2.89 & 33.74 $\pm$ 2.82 & 30.62 $\pm$ 3.68 \\
		\hline \hline
	\end{tabular}}
	\vspace{1mm}
	\\
	{The best performance results are indicated by boldface numbers.}
	\label{tab_2}
\end{table*}

\begin{figure*}[t]
	\centering
	\includegraphics[trim = 2mm 2mm 0mm 2mm, clip, scale=0.818]{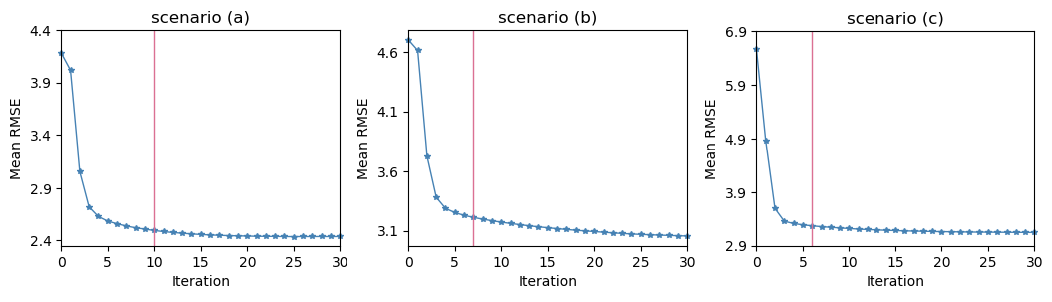}
	\caption{RMSE convergence mean curves (blue) of our method with the CAVE dataset and blurring scenarios (a), (b) and (c). Red lines represent the iteration number given by the proposed stopping criterion.}
	\label{fig:mrmse}
\end{figure*}

\begin{figure*}[t]
	\centering
	\includegraphics[trim = 2mm 2mm 0mm 2mm, clip, scale=0.80]{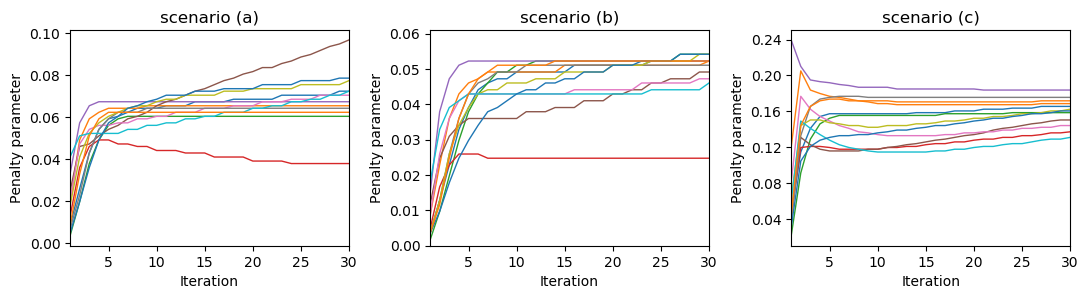}
	\caption{Estimated penalty parameters $\rho_k$ as a function of iteration index $k$, for different images of the CAVE dataset and blurring scenarios (a), (b) and (c). Lines with different colors refer to different test images. }
	\label{fig:rho}
\end{figure*}

\subsection{Performance evaluation on simulated data}
We start validating the tuning-free scheme with the CAVE dataset by demonstrating its effectiveness in terms of HSI deconvolution performance over the other methods. 

Table~\ref{tab_1} reports the average values and standard deviations of RMSE, PSNR, SSIM and ERGAS. For all blurring scenarios, one can observe that our method outperformed all competing methods in terms of performance and robustness. 
For quality comparison, consider scenario (a) for example. {Fig.}~\ref{fig:cave} provides the blurred image, deblurred images, ground truth of \emph{real and fake peppers} (first row) and \emph{superballs} (second row) from the CAVE dataset. Visually, our method provides more details, including sharper edges and more vivid gloss. This confirms the effectiveness of the proposed method in recovering the spatial information of the latent clear HSIs. {To further evaluate the robustness of the proposed method, consider scenario (a) for example. We set the noise level from 0.01 to 0.05 at an interval of 0.01 to generate varying noise interruptions added to input images.  
In Table~\ref{tab_noise}, it can be seen that the performance of all methods deteriorates with the increase in noise levels. However, the proposed method still provides the best quantitative results with all different noise interruptions.}

We now evaluate the proposed method on remotely sensed data: the Chikusei dataset. This dataset, with more spectral bands, allows to analyze how our method exploits spectral information. The mean and variance of the numerical results for all methods in 6 blurring scenarios are provided in Table~\ref{tab_2}. It can be observed that the quantitative metrics of our method surpass the other competing methods in most cases. 
{Fig.}~\ref{fig:chikusei} displays the visual results. As can be seen, the proposed method provides results with clearer and sharper visual effects compared to the other methods. This illustrates the superiority of our method in recovering the latent HSIs with more spectral bands.

\subsection{Convergence illustration}
\label{convergence}

%  \begin{table}[tp]
% 	\centering
% 	\caption{{Average RMSE, PSNR, SSIM and ERGAS of the proposed method with different parameter setting policies for the CAVE dataset with the blurring scenario (a).}} 
% 	\vspace{2mm}
% 	{	\begin{tabular}{c|cccc}
% 			\hline \hline
% 			Policies      &  RMSE                 &   PSNR         &  SSIM      &      ERGAS              \\ \hline \hline
% 			\multirow{4}{*}{\begin{tabular}[c]{@{}c@{}} Iter $1$ \\ Iter $15$ \\ Iter $30$  \\ Proposed  \end{tabular}} &
% 			\multirow{4}{*}{\begin{tabular}[c]{@{}c@{}} 3.531 \\ 3.070 \\ 3.062 \\ 3.132  \end{tabular}} & \multirow{4}{*}{\begin{tabular}[c]{@{}c@{}} 38.452 \\39.443 \\ 39.462 \\ 39.252  \end{tabular}} & \multirow{4}{*}{\begin{tabular}[c]{@{}c@{}} 0.9158\\0.9494\\ 0.9496 \\ 0.9493  \end{tabular}} & 
% 			\multirow{4}{*}{\begin{tabular}[c]{@{}c@{}} 13.91\\12.66\\12.65\\13.01 \end{tabular}}  \\ 
% 			&           &      &    &    \\ 
% 			&           &      &    &    \\
% 			&           &      &    &    \\ \hline \hline
% 	\end{tabular}	}\\\vspace{1mm}
% 	{The best results are indicated by boldface numbers.}
% 	\label{tab_rho}
% \end{table}

In many PnP algorithms for inverse imaging problems, the ADMM is widely used as a variable splitting technique. In some works, the convergence of PnP schemes based on some linear denoisers, including Non-Local Means (NLM)~\cite{sreehari2016plug} and Gaussian Mixture Model (GMM)~\cite{teodoro2017scene}, has been proved theoretically. 
%  However, in our work, we use a deep learning based denoiser as a black-box to replace the explicitly solving the second subproblem~\eqref{eq:stepb}, and thus the convergence of the proposed framework is complex to analysis.
It is difficult if not impossible to prove the convergence of our method as the B3DDN denoiser involves amounts of non-linear operators. In practice, however, as illustrated below, we observed that the proposed deconvolution framework shows good convergence behaviour.

Figure~\ref{fig:mrmse} provides the mean RMSE curves of our algorithm
obtained for the CAVE dataset in the case of scenarios (a), (b) and (c).
It can be observed that the algorithm, even with its nonlinear B3DDN denoiser, exhibits a stable and robust convergence behaviour independently of the blurring kernel and noise level.
Moreover, a low mean RMSE value was reached after few iterations, which indicates that early stopping can be considered to limit computation time.

\subsection{Behavior with respect to PnP internal parameter estimation}
\label{behavior}
Deep priors that capture both the spatial context and spectral correlations of the latent clean HSIs mainly contribute to the effectiveness of our method. But the internal parameter setting procedure and the stopping criterion also play a crucial role in achieving satisfactory performance by yielding a good balance with the contribution of deep priors.
In contrast, observe that the automatic setting of the regularization parameters is not implemented by the other competing methods during test.

{Fig.}~\ref{fig:rho} shows how the penalty parameter varies along with the iterations, for different images of the CAVE dataset, and for scenarios (a), (b), and (c). According to the PnP principle, the estimated noise level $\sigma_k$ is assumed to decrease along with
the iterations, as the reconstructed image converges to a desired point. Therefore, the penalty parameter $\rho_k = \lambda / \sigma_k^2$ is expected to increase~\cite{zhang2021plug}. As can be seen on Fig.~\ref{fig:rho}, parameter $\rho$ changes coincide with this trend for almost all test images.  {Fig.}~\ref{fig:mrmse} shows the number of iterations $K$ for scenarios (a), (b) and (c). It can be observed that our stopping criterion automatically interrupts the PnP algorithm when it has nearly converged, which contributes to save computation time.

\begin{figure*}[t]
	\centering
	\includegraphics[trim = 0mm 0mm 0mm 0mm, clip, scale=0.71]{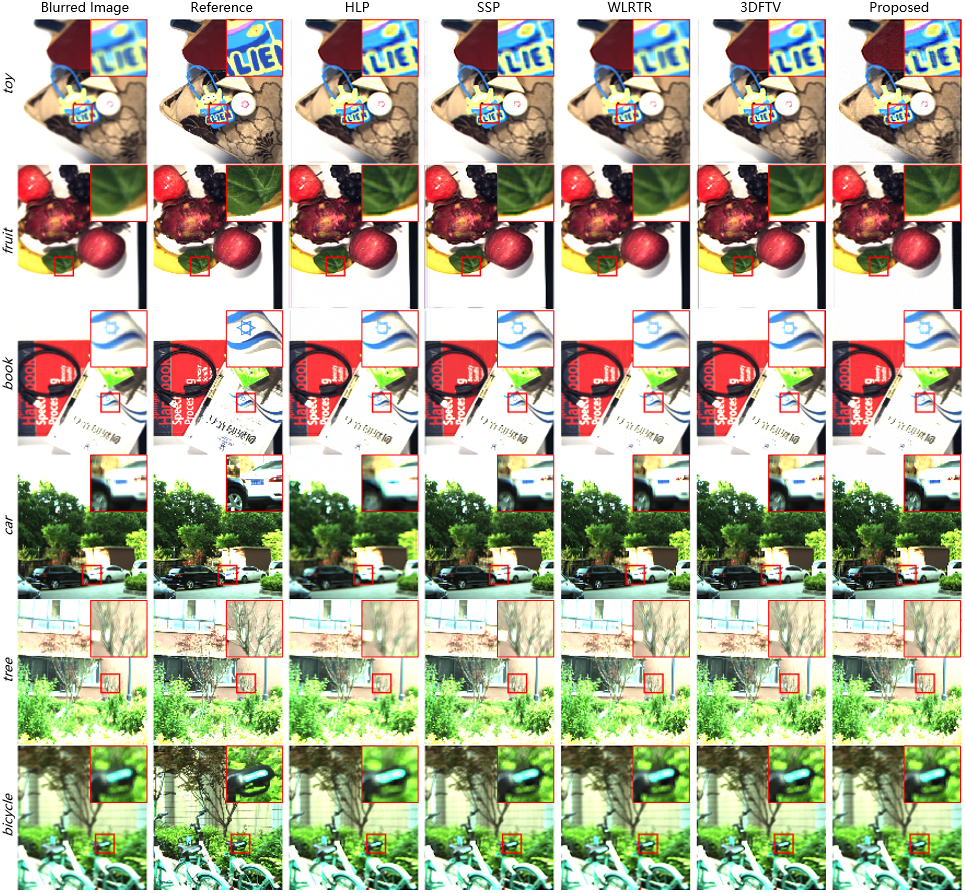}
	\caption{{Blurred images, reference images and visual results for all methods on the real-world dataset. The false color images were generated for clear visualization with the 38th, 24th and 10th channels used for red, green and blue, respectively.}}
	%\cblue{[Jie: The prior information was learnt from CAVE which mainly consists of quotient objects. However this image includes objects that are somewhat different from CAVE, e.g., the circuit board. This might cause performance loss. If necessary, we may capture another HSI.]}}
	\label{fig:real}
\end{figure*}

\begin{figure}[t]
	\centering
	 \begin{minipage}{0.48\linewidth}
	\includegraphics[scale=0.1]{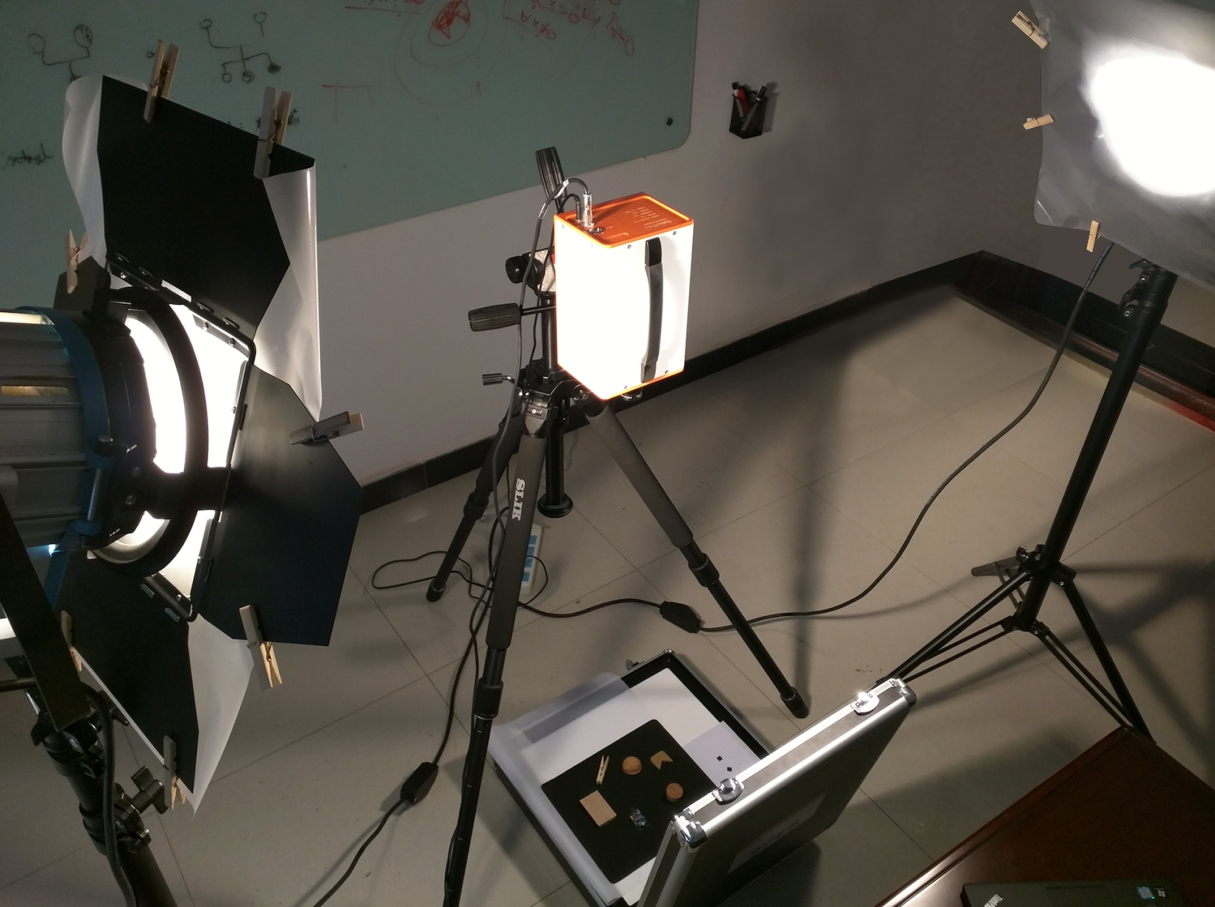}
	    \end{minipage}
    \begin{minipage}{0.48\linewidth}
    \centering
    \includegraphics[scale=0.03327]{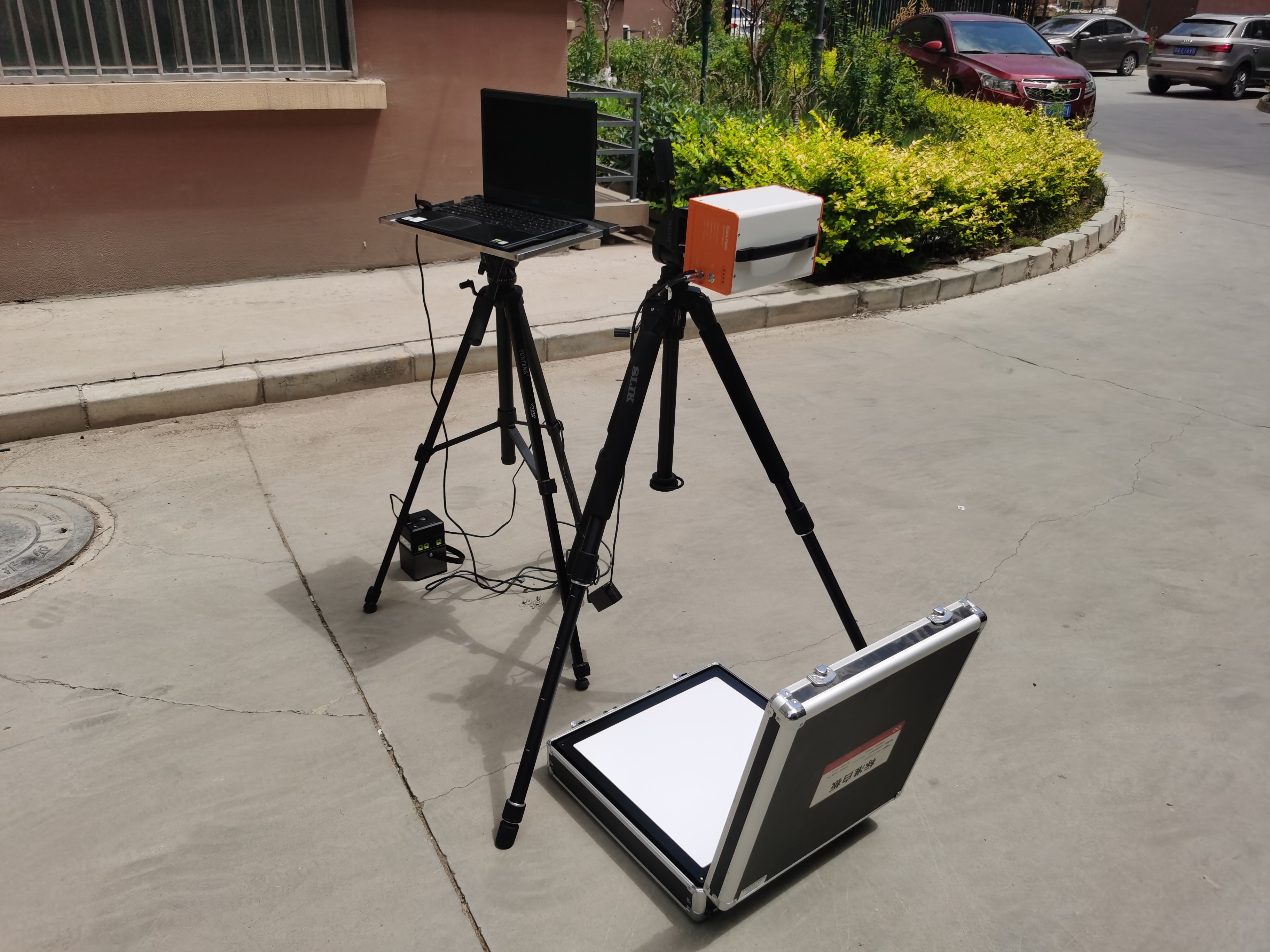}
    \end{minipage}
	\caption{Indoor (left) and outdoor (right) experimental setups for collecting real data.}
	\label{fig:equipment}
\end{figure}

\begin{figure}[t]
	\centering
	\includegraphics[trim = 0mm 1mm 0mm 1mm, clip, scale=0.84]{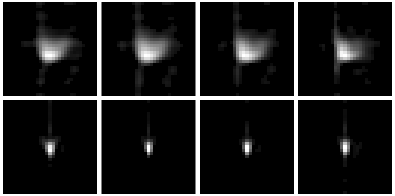}
	\caption{Estimated blurring kernels in the 10th, 20th, 30th and 40th channels of the blurred images \emph{fruit} (first row) and \emph{bicycle} (second row) of the real-world dataset.}
	\label{fig:kernel_real}
\end{figure}

\subsection{Performance evaluation on real-world data}
\label{subsec:real}
To validate the effectiveness of our method in real-world conditions, we collected six unfocused HSIs and the corresponding focused images for different indoor and outdoor scenes.
Specifically, as illustrated in Fig.~\ref{fig:equipment}, the HSIs of the indoor scenes were recorded under controlled illuminations while the outdoor HSIs were captured under normal daylight illumination. To fully capture the complex blurs caused by the imaging system, our dataset was elaborated to address hyperspectral deconvolution problem with respect to defocus. In particular, blurred images were obtained by making the camera out of focus while clear references were also captured by
focusing the camera. We captured these images with the GaiaField systems (see details in~\cite{zhao2019laboratory}) of our laboratory at Northwestern Polytechnical University. The GaiaField (Jiangsu Dualix Spectral Image Technology Co. Ltd., GaiaField-V10) is a push-broom imaging spectrometer with an HSIA-OL50 lens, covering the visible and NIR wavelengths ranging from 373.70 to 1000.90 nm, with a spectral resolution of 4.6 nm (129 channels in total). The spatial resolution of the images is $780\times 696$ pixels. 

For all acquired images, we conducted a pre-processing procedure as described in~\cite{simoes2015HySure}. First, we removed over-noisy and over-exposed bands. We got 45 exploitable bands, which were normalized such that the 0.999 intensity quantile corresponded to the value 1. Then, all HSIs were denoised using the approach described in~\cite{roger1996denoisingHSI} to enhance images. Blurred images and their clear counterparts are illustrated in the first and second columns of Fig.~\ref{fig:real}, respectively. Note that these image pairs are not strictly aligned due to multiple factors affecting the camera mounting. The clear images were used for visual comparisons only.
The blurring kernel in each channel was estimated using the method described in~\cite{krishnan2011blind}. For illustration purpose, {Fig.}~\ref{fig:kernel_real} shows the kernels in the 10th, 20th, 30th and 40th channels of the blurred images \emph{fruit} and \emph{bicycle}. For all experiments, we added an i.i.d. Gaussian noise to the blurred images, with a signal-to-noise ratio (SNR) set to 40~dB.

In real-world HSI deconvolution scenarios, no ground truth is available for training the B3DDN. Benefiting from the flexibility of the B3DDN in denoising HSIs of various origins with distinct numbers of spectral bands, in this experiment we used the network parameters $\Theta$ learned with the CAVE dataset (31 spectral bands). {Fig.}~\ref{fig:real} shows the deblurred images obtained with all the competing algorithms, from columns 3 to {7}. It can be seen that our method still performed better, or similarly, in recovering details compared to HLP, SSP, WLRTR, {and 3DFTV,} though all competing methods only achieved limited performance probably due to deviations in estimating kernels. 
This demonstrates the applicability of our method in real-world scenarios, as well as the necessity of further investigating blind hyperspectral deconvolution algorithms. 

{Finally, we conducted the experiment for evaluating the running time using the blurred image \emph{fruit} from our real-world dataset. All the baselines were implemented using MATLAB while our method was carried out using Python. We conducted all the experiments on a server with
Intel Xeon Gold 6152 CPU, 512-GB random access memory and NVIDIA Tesla P40 GPU. 
Time consuming of all the compared methods is shown in
Table~\ref{tab:time}.  
It can be observed that our method achieves most competitive deconvolution results with relatively less computation time.}

\begin{table}[t]
\footnotesize \centering
\caption{Time consuming of the compared methods for the blurred image \emph{fruit} of the real-world dataset.}
\renewcommand{\arraystretch}{1.5}
{\begin{tabular}{c|ccccc}
\hline\hline
        & HLP & SSP & WLRTR & 3DFTV & Ours \\ \hline
Time (sec) & 9.7  & 622.5  & 10501.2  &  6044.4 &  4280.6 \\
%Time (sec) & 8.3  & 485.4  & 5337.5  & 4741.5  & 4892.2  \\
\hline\hline
\end{tabular}}
\label{tab:time}
\end{table}

\section{Conclusion}
\label{sec:conclusion}
In this paper we presented a tuning-free HSI deconvolution method based on the PnP framework. Instead of using handcrafted priors, we designed a blind B3DDN denoiser based on deep learning to learn the spectral-spatial information of hyperspectral images from data and plugged it into an ADMM optimizer. The internal parameters were automatically estimated by a measure of 3D residual whiteness and learned by the B3DDN during iterations. Experimental results demonstrated that the proposed method can not only effectively handle various simulated blurring settings but can also be applied to real-world scenarios. In the future, we will address blind HSI deconvolution {and computational cost reduction} to further enhance the applicability of our method in real-world scenarios.

% In the future, we will extend the proposed HSI deconvolution method in two directions. On the one hand, blind deconvolution with joint blurring kernel estimation process will be addressed. On the other hand, computational cost reduction during the deconvolution procedure will be investigated. These works will further enhance the applicability of our method in real-world scenarios.

% \appendices 

% \section{Proof of Theorem 2}
% \label{appendix:b}

%\newpage


\bibliographystyle{IEEEtran}
\bibliography{mybib}

\end{document}